\documentclass[12pt]{iopart}
\pdfoutput=1
\usepackage{iopams}
\usepackage{ulem}
 
\expandafter\let\csname equation*\endcsname\relax
\expandafter\let\csname endequation*\endcsname\relax
\usepackage{amsmath,amssymb,mathtools}

\usepackage[colorlinks,linkcolor=blue,urlcolor=blue,citecolor=blue]{hyperref}

\def\be{\begin{equation}}
\def\ee{\end{equation}}
\def\bea{\begin{eqnarray}}
\def\eea{\end{eqnarray}}

\usepackage{xcolor}
\definecolor{dgreen}{rgb}{0,0.7,0}

\begin{document}

\title{Stochastic resetting with stochastic returns using external trap}

\author{Deepak Gupta$^{1}$, Carlos A Plata$^{1}$, Anupam Kundu$^{2}$ and Arnab Pal$^{3}$\footnote{All the authors have  contributed equally to this work.}}
\address{ \textit{$^{1}$Dipartimento di Fisica `G. Galilei', INFN, Universit\`a di Padova, Via Marzolo 8, 35131 Padova, Italy}}
\address{\textit{$^{2}$International Centre for Theoretical Sciences, Tata Institute of Fundamental Research, Bengaluru 560089, India}}
\address{\textit{$^{3}$School of Chemistry, The Center for Physics and Chemistry of Living Systems, Tel Aviv University, Tel Aviv 6997801, Israel}}

\begin{abstract}
In the past few years, stochastic resetting has become a subject of immense interest. Most of the theoretical studies so far focused on instantaneous resetting which is, however, a major impediment to practical realisation or experimental verification in the field. This is because in the real world, taking a particle from one place to another requires finite time and thus a generalization of the existing theory to incorporate non-instantaneous resetting is very much in need. In this paper, we propose a method of resetting which involves non-instantaneous returns facilitated by an external confining trap potential $U(x)$ centered at the resetting location. We consider a Brownian particle that starts its random motion from the origin. Upon resetting, the trap is switched on and the particle starts experiencing a force towards the center of the trap which drives it to return to the origin. The return phase ends when the particle makes a \textit{first passage} to this center. We develop a general framework to study such a set up. Importantly, we observe that the system reaches a non-equilibrium steady state which we analyze in full details for two choices of $U(x)$, namely, (i) linear and (ii) harmonic. Finally, we perform numerical simulations and find an excellent agreement with the theory. The general formalism developed here can be applied to more realistic return protocols opening up a panorama of possibilities for further theoretical and experimental applications.
\end{abstract}

\section{Introduction}
\noindent
The subject of resetting or restart has recently taken the center stage due to its myriad of applications in statistical physics \cite{Restart1,Restart2,Kirone,transport1,transport2,Durang,Pal-potential,Pal-time-dep,VV,SRRW,underdamped,local-time,restart_conc17,magnetic,ca-sa}, in chemical
and biological physics \cite{ReuveniEnzyme1,ReuveniEnzyme3,bio,RT}, in computer science \cite{Luby,Montanari}, and in search theory \cite{Restart-Search1,Restart-Search2,Restart-Search3,Chechkin,HRS} (see \cite{Review} for a broad review of the topic). Resetting is a simple mechanism where one intermittently stops an ongoing dynamics only to start over again. Despite its simple enough implementation, there has been a surge of rigorous theoretical investigations to understand the effect of resetting on generic stochastic processes.  In this context, there are many different aspects that have been studied under the umbrella of non-equilibrium phenomena including stationary states and first passage problems. As an example, consider the paradigmatic model of diffusion in the presence of resetting \cite{Restart1}. In this model, it has been shown that resetting renders a non-equilibrium steady state even though the underlying process i.e., a simple diffusion without resetting is non-stationary. Similar observation along with interesting relaxation phenomena were also made in other stochastic processes, namely anomalous diffusion \cite{anamolous-1,anamolous-3}, fractional Brownian motion \cite{FBM}, scaled Brownian motion \cite{scaled}, continuous time random walk \cite{CTRW1,CTRW2}, L\'evy flights \cite{Levy-flight}, and extended systems with many degrees of freedom \cite{KPZ,SEP,ASEP,SEP2}. While appropriate resetting mechanism can possibly make a non-stationary process stationary, in the context of first passage time problems it has been shown that resetting mechanisms have dramatic consequences
on the completion times of stochastic processes. In particular, repeated resetting can facilitate the completion of an arbitrary complex process \cite{Restart1,Restart2,Pal-time-dep,Restart-Search1,Chechkin,Review,ReuveniPRL,PalReuveniPRL,branching,Belan,Landau,Peclet,Gupta,log} and this has lead to the major challenging questions such as improvement  of the search protocols using resetting \cite{Pal-time-dep,PalReuveniPRL}, existence of an optimal resetting protocol \cite{Pal-time-dep,Chechkin,PalReuveniPRL} and the universal properties associated to such optimal protocol \cite{ReuveniPRL,Landau,ORR}.

Most of the  studies on stochastic processes with resetting unequivocally
assume the resetting mechanism to be
instantaneous. For example, consider the paradigmatic model of diffusion with resetting where the particle is relocated to the initial configuration in zero time \cite{Restart1,Restart2,Review}. However, these underlying assumptions are often over simplified since in reality the particle will require certain amount of time to return. Thus, new formulations need to be more realistic and physically amenable to include such possibilities into account. Moreover, physical intuition tells us that the return time of the particle will typically depend on its current location, and also on other environmental parameters during its return. A few initial attempts in this direction have recently been proposed in \cite{HRS,Pal-19-06} and other consecutive studies \cite{invariance1,invariance2,return3,return4} where the returns were considered to be non-instantaneous 
{\it but deterministic}. Examples include, return by a constant or spatially homogeneous velocity \cite{invariance1,invariance2,return3,return4}, return in certain time \cite{refractory} or by means of other deterministic motions \cite{return3}. Only recently, experiments using optical tweezers have been performed to realize such protocols \cite{expt,expt2}. 
\begin{figure}
    \centering
     \includegraphics[width=12cm]{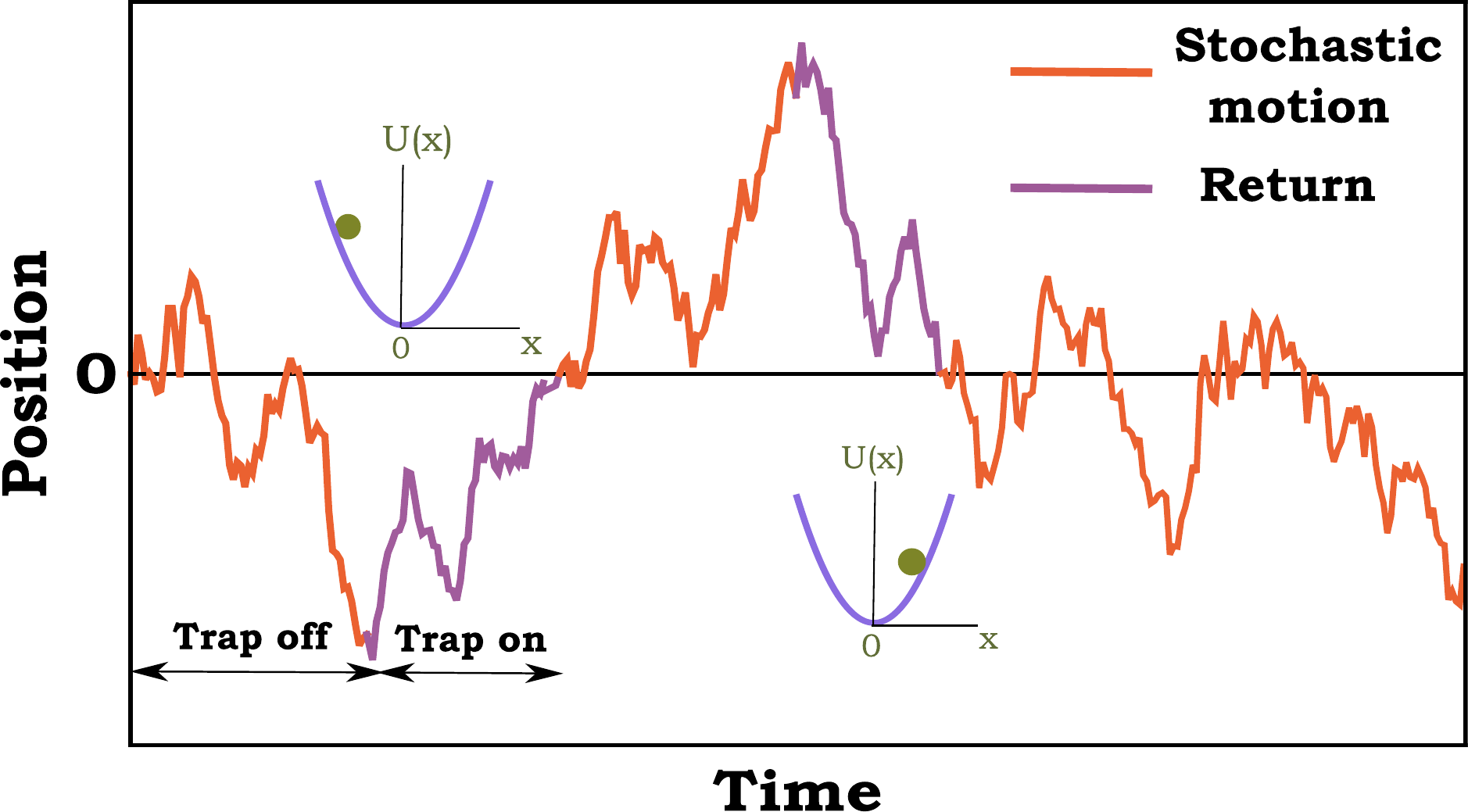}
   \caption{Schematic trajectory of a particle undergoing a stochastic motion phase (e.g., free diffusion) followed up by a return phase modulated by an optical trap which is centered at the origin. The trap creates an attractive potential $U(x)$ around its center. The trap is activated at the time of resetting, which occurs at random with rate $r$, and kept `on' until the particle returns to the origin for the \textit{first time}. Once the particle 
   hits the origin, the potential is switched `off' and the particle resumes its stochastic motion phase.}
\label{trajectory}
\end{figure}

We consider the following protocol: assume that a particle starts a random motion (phase I) from the origin at time zero. It explores around until a resetting occurs after some random interval of time and consequently the particle is required to return to the origin (for simplicity we set the initial position same as the resetting position). This phase is called the return phase (phase II). In this phase the motion of the particle is manipulated externally by turning an external confining potential on. Such confining potential can be achieved in experiments using the optical trap technique \cite{optical-trap}. The trap (often modelled as harmonic potential in theory) is set up so that its minimum is at the origin and it is switched on only during phase II. When active (i.e., on), the trap generates an attractive force field towards its center (the origin), and hence the particle will diffuse downhill (phase II - return). We observe the motion until the particle hits the origin for the first time. At this very moment, the trap is switched off and the particle resumes its exploring phase motion i.e., phase I motion (see \fref{trajectory} for a typical trajectory). It is easy to realise that the compound process is recurrent and in the long time the particle will reach a non-equilibrium steady state.  In this paper, we are interested in the distribution of the position of the particle in this steady state.

The remainder of the paper is organized as follows. We provide the details of our set-up in \sref{sec:Setup}. In the next section (\sref{sec:Gen_Th}) we develop a general theory which characterizes such set up within a broad framework. Sec. \ref{sec:st_st} is devoted to the analysis of the non-equilibrium stationary state. We then apply our framework to some realistic situations, and derive exact results in \sref{sec:ex}. We conclude the paper in \sref{sec:concl} with discussions and possible future directions.

\section{Set up}
\label{sec:Setup}
We consider a set up in one-dimension in which a particle starts to evolve from the origin at time zero. Motion of the particle is governed by a two-phase compound process. In the first phase, the particle performs motion according to some
stochastic process. In the second phase it returns to its initial position with the aid of an external force exerted by the optical trap. A typical trajectory of such a compound process is shown in \fref{trajectory}. In the following, we describe each of the phases in details.

\textit{Phase I - Stochastic motion and resetting}:
We assume that the first phase of stochastic motion of the particle can be described by a Fokker-Planck equation \bea
\partial_t \rho_0(x,t)=\mathcal{L}_0 \rho_0(t)~,
\label{underlying}
\eea
where $\mathcal{L}_0$ is the generator of the stochastic process and $\rho_0(x,t)$ is the propagator that describes the particle's motion when observed in the absence of resetting. To understand the mechanism of stochastic resetting let us briefly discuss the case of instantaneous resetting in which at any small interval $dt$, the particle is instantaneously teleported to the origin with probability $rdt$. In other words, the
waiting times between resetting events is drawn from an exponential distribution $f(t)=re^{-rt}$. The corresponding master equation reads \cite{Restart1,Restart2,Pal-potential}
\bea
\partial_t\rho_{\text{inst}}(x,t)=\mathcal{L}_0 \rho_{\text{inst}}(x,t)-r \rho_{\text{inst}}(x,t)+r \delta(x)~,
\label{inst-propagator}
\eea
where $\rho_{\text{inst}}(x,t)$ is the propagator for such a process. Stochastic processes with instantaneous resetting have been studied in great details. Here we consider the situation where the particle takes non-zero time to return to the origin. This will be discussed in the following.

\textit{Phase II - Return:} 
As done in the context of instantaneous resetting,  after a random time interval $t$ drawn from the exponential distribution $f(t)$, motion of the particle changes from phase I to phase II. In this phase, the particle is facilitated to return to the origin by switching on an external confining 
potential $U(x)$. The center of the trap is situated at the resetting location 
(same as the initial position and set at the origin). Thus, the particle will diffuse in the force field $F(x)$ generated by the potential $U(x)$ and eventually will reach at the minimum of the trap. The first time the particle reaches to the center (origin), the trap is switched off and this marks the end of phase II. Upon return to the origin, the particle momentarily restarts its motion in phase I. Therefore, we have a compound event or a trial which comprises of the stochastic motion in phase I until
the resetting epoch followed by a first passage to the center of the trap in phase II. Each trial repeats by itself along a long trajectory and is essentially a renewal process. We mention that similar switching `on' and `off' of external potentials is also used in ratchet problems \cite{ratchet-rev}, \textit{albeit} with an important difference. There, switching between `on' and `off' states of the potential are usually controlled by external agents, whereas in our set up this depends entirely on the motion of the particle in the potential in phase II.
 In the next section we provide a general formalism of stochastic resetting under stochastic returns.

\section{General theory}
\label{sec:Gen_Th}
In this section, we will construct a set of master equations, similar to \eref{inst-propagator}, to describe the full dynamics of our system. 

\subsection{Dynamical equations for motion and return density} 
We assume that the particle, starting from the origin, performs simple diffusion (with diffusion constant $D$) in phase I and upon resetting it returns to the origin via diffusion in the force field $F(x)$ with a different diffusion constant $D_R$. Although in any experimental set up it is reasonable to consider $D_R=D$, we, however, proceed with $D_R \neq D$ to keep our framework more general.

Let, $\rho_M(x,t)$ and $\rho_R(x,t)$ be the probability densities for the position of the particle at time $t$ during the stochastic motion (M - phase I) and the return (R - phase II) phases, respectively.  Thus the full density $\rho(x,t)$ of the combined process  is given by 
\bea
\rho(x,t)=\rho_M(x,t)+\rho_R(x,t)~,
\label{rho-tot}
\eea
which consists of two parts with respective contributions from each phase. It is evident that $\rho_M(x,t)$ and $\rho_R(x,t)$ are not individually normalized as their sum is the total probability density $\rho(x,t)$ which is normalized to one. The probabilities to find the particle in the stochastic motion (phase I) and return phase (phase II) are respectively given by
\begin{align}
\begin{split}
    p_M(t) \equiv \text{Prob}(\text{\text{Phase I}}) &= \int\limits_{-\infty}^{\infty} dx~\rho_M(x,t), 
    \\
    p_R(t) \equiv \text{Prob}(\text{\text{Phase II}}) &= \int\limits_{-\infty}^{\infty} dx~\rho_R(x,t),
    \label{pa-pb}
\end{split}
\end{align}
 where $p_M(t)+p_R(t)=1$ at all times with the initial condition $p_M(0)=1$.
 
 We now derive equations for $\rho_M(x,t)$ and $\rho_R(x,t)$, and thus for the propagator $\rho(x,t)$ which describes our process. We first note that
the probability current at $x$ due to returning particles is $J_R(x,t)=F(x) \rho_R(x,t)-D_R \frac{\partial \rho_R(x,t)}{\partial x}$. Note that, without loss of generality, we set the friction coefficient to be unity through out the paper.   On the other hand, the probability flux at $x$ due to particles which switch from the stochastic motion in phase I to the return phase is given by $r\rho_M(x,t)$. Adding these two possibilities gives 
\bea
    \partial_t \rho_R(x,t) &=&
    -\partial_x J_R(x,t)+ r\rho_M(x,t) -\delta(x)\left[J_R(0^-,t)-J_R(0^+,t)\right]~, 
    \label{rhoR-eqn}
\eea
where the last term on the right hand side serves as a sink and accounts for the fact that the returning particles switch to stochastic motion in phase I as soon as they arrive at the origin. Note that the current at the origin gets contributions from the incoming return particles from both sides, and these two contributions are not necessarily equal in magnitude (in case of asymmetric potential applied in the return phase). Thus in this case $J_R(0^+,t)$ [$J_R(0^-,t)$] is simply the current flowing into the origin from the positive (negative) side on the real axis. We define the net current at the origin as the total probability flux arriving at $x=0$ from both sides, that is,
$
\overline{J}_R(0,t)=J_R(0^-,t)-J_R(0^+,t)
$. Thus, we can rewrite \eref{rhoR-eqn} as
\bea
\partial_t \rho_R(x,t)&=-\partial_x J_R(x,t)+r \rho_M(x,t)- \delta(x) \overline{J}_R(0,t)
\label{rhoR-eqn2}
\eea
We now turn our attention to the time evolution of the position distribution  $\rho_M(x,t)$ in phase I. During this phase, the particles will evolve in $dt$ time interval according to the generator of the process $\mathcal{L}_0$ with probability $1-rdt$. With the complementary probability, the particles will switch from this phase to the return phase. Accounting for all the contributions, we arrive at the following equation
\bea
    \partial_t \rho_M(x,t) = \mathcal{L}_0 \rho_M(x,t)-r \rho_M(x,t)  + \delta(x) \overline{J}_R(0,t),
    \label{rhoD-eqn}
\eea
where the second term on the RHS indicates the outward probability from the stochastic motion (phase I). The third term on the RHS acts as a source for the returning particles in phase II which switch to stochastic motion in phase I upon arrival at the origin. Thus, \eref{rhoR-eqn2} and \eref{rhoD-eqn} together with the initial conditions $\rho_R(x,0)=0$ and $\rho_M(x,0)=\delta(x)$ describe the complete motion of the particle. 

\subsection{Boundary and matching conditions on the densities}
\label{bcss}
To solve the master equations in phase I and II, one needs to supplement them with appropriate boundary conditions which we discuss in the following. We start by decomposing the return phase density in the following way
\bea
   \rho_R(x,t)=\rho_R^-(x,t) \theta(-x)+\rho_R^+(x,t) \theta(x)~,
\eea
where $\rho_R^{\pm}(x,t)$ correspond to the phase space densities conditioned on the fact that the particles started their return phase motion from $x>0~ (x<0)$ respectively at the end of phase I and $\theta(x)$ stands for the Heaviside step function. This decomposition is natural since the returning particles can either be in the positive or negative $x$ in each trial. Thus, switching between them is possible only via returns to the origin followed by the phase I. Essentially this implies that $\rho_R^{\pm}(x,t)$-s are not directly coupled, but rather coupled through $\rho_M(x,t)$. The returning particles, upon reaching at the origin for the first time, switch to phase I. As a result, the origin effectively acts as an absorbing boundary for the return probabilities. Hence, we should have
\bea
\rho_R^{\pm}(0,t)=0~.
\label{rhoR-abs}
\eea
It should be stressed again that there is no physical absorbing boundary at the origin, but the above conditions are invoked to implement the termination of return phase II at the origin.
On the other hand the particles can not get accumulated at the infinity in finite time for motions in both phases I and II. As a result, we have
\bea
\lim_{x \to  \pm \infty} \rho_R^{\pm}(x,t)=0, \qquad \lim_{x \to  \pm \infty} \rho_M(x,t) =0.
\label{rho-infinity}
\eea
Equivalently there can not be any particle current at $x= \pm \infty$ from phases I and II, which implies the following natural boundary conditions
\bea
\lim_{x\to \pm \infty} J_R(x,t)=0, \qquad \lim_{x\to \pm \infty} J_M(x,t)=0.
\label{JR01}
\eea
These conditions can also be argued from the more natural boundary conditions i.e., $\lim \limits_{x \to \pm \infty}J_{\text{tot}}(x,t) =0$, where $J_{\text{tot}}(x,t)$ is the total particle current which satisfies the full continuity equation $\partial_t \rho(x,t)+\partial_xJ_{\text{tot}}(x,t)=0$. The above equation is obtained by simply adding \eref{rhoR-eqn2} and \eref{rhoD-eqn}. We can therefore write $J_{\text{tot}}(x,t)=J_R(x,t)+J_M(x,t),$ where $J_{M}(x,t)$ can be understood as the particle current in the phase I motion. A careful observation of \eref{rhoD-eqn} tells us that $J_M(x,t)$ is not sensitive to the arbitrary choices of the return dynamics in phase II. As a result, the $x$ dependence of $J_M(x,t)$ is completely independent from that of $J_R(x,t)$. Moreover, one also has the freedom to choose two independent dynamics for the positive and negative $x$ in phase II. Thus, $x$ dependence of $J_R(x,t)$ at $x>0$ is also independent from the $x$ dependence of $J_R(x,t)$ at $x<0$. Now if $J_{\text{tot}}(x,t)$ vanishes as $x \to \pm \infty$, the individual currents $J_M(x,t)$ and $J_R(x,t)$ also must go to zero as $x \to \pm \infty$. This provides another justification of the boundary conditions in \eref{JR01}.

Finally, we note that $\rho_M(x,t)$ is continuous across the origin implying
\begin{align}
\rho_M(0^{+},t)=\rho_M(0^{-},t)~,
\label{matching}
\end{align} 
as a matching condition for solving the density in phase I. Since the boundary conditions in Eqs. (\ref{rhoR-abs}) - (\ref{matching}) are valid at all times, they automatically translate to the steady state.

\section{Non-equilibrium steady state}
\label{sec:st_st}
Equations (\ref{rhoR-eqn2}) and (\ref{rhoD-eqn}) along with the initial and boundary conditions stated above constitute the central equations that describe a full, time dependent, description of stochastic motion with resetting and space time coupled stochastic returns to the origin. However, solving these equations in full generality seems quite challenging and from now on, we focus only on the steady-state properties of our process.

\subsection{Steady state density in phase I}
To compute the density in phase I, we first show that the steady state equation (obtained by setting $\partial_t \rho_M(x,t)=0$ in \eref{rhoD-eqn}) for $\rho_M(x)$ can be cast into the form satisfied by the steady state density $\rho_{\text{inst}}(x)$ for instantaneous resetting. This can be done by observing that $\overline{\rho}_M(x)=\frac{r}{\overline{J}_R(0)}\rho_M(x)$ satisfies 
\bea
\mathcal{L}_0 \overline{\rho}_M(x) -r\overline{\rho}_M(x) +r \delta(x)=0~,
\eea
which has the same form as \eref{inst-propagator} in steady state. Here  $\overline{J}_R(0)=\lim_{t \to \infty} \overline{J}_R(0,t)$ is the steady state current to the origin. Hence we have 
\bea
\rho_M(x)=\frac{\overline{J}_R(0)}{r} \rho_{\text{inst}}(x)~,
\label{rhoM-0}
\eea
which directly implies (after integrating both sides of the above equation over the real axis)
\bea
\overline{J}_R(0)=rp_M~,
\label{JR0}
\eea
where $p_M$ (now time independent) is the stationary probability that the particle is in  phase I. Substituting this relation in \eref{rhoM-0}, we have
\bea
\rho_M(x)=p_M\rho_{\text{inst}}(x)~.
\label{equivalence1}
\eea
It is important to point out that $\rho_M(x)=p_M \rho(x|\text{phase I})$ where $\rho(x|\text{phase I})$ is the conditional probability density to find the particle at a position $x$ given that it is in phase I. Thus, \eref{equivalence1} simply asserts $\rho(x|\text{phase I})=\rho_{\text{inst}}(x)$ which will prove to be a useful relation later. In the next section, we discuss how to obtain the steady state density in phase II.

\subsection{Steady state density in return phase}
To compute the steady state density in return phase, we proceed as before and set $\partial_t \rho_R(x,t)=0$ in \eref{rhoR-eqn2}. Further we substitute $\overline{J}_R(0)$ from \eref{JR0} into \eref{rhoR-eqn2} to obtain
\bea
D_R \frac{\partial^2  \rho_R(x)}{\partial x^2} -\frac{\partial}{\partial x} \left[ F(x) \rho_R(x) \right] +r \rho_M(x)-rp_M\delta(x) =0~,~\hspace{0.3cm}
\label{rho-R-full}
\eea
which can, in principle, be solved for a given $F(x)$.
However, as we have shown earlier $\rho_R(x)$ can again be split into the density in positive and negative axis so that
\begin{equation}
 \rho_R(x)=\rho_R^-(x) \theta(-x)+\rho_R^+(x) \theta(x)~,
 \label{jointrho-R}
\end{equation}
where the  probabilities on positive and negative $x$, individually satisfy 
\bea
D_R \frac{\partial^2  \rho^{\pm}_R(x)}{\partial x^2} -\frac{\partial}{\partial x} \left[ F(x) \rho^{\pm}_R(x) \right] +r \rho_M(x)=0.
\label{conditional}
\eea
Applying the boundary conditions as mentioned in \sref{bcss} and utilizing the form for $\rho_M(x)$, we note that the solutions $ \rho_R(x)$ can be expressed in terms of $p_M$. Hence, the remaining task at our disposal is to evaluate $p_M$.

\subsection{Interpretation of $p_M$}
\label{pm}
There are two different ways to solve for $p_M$. In the first method, we simply use the normalization condition
for the total probability density i.e.,
\bea
\int \limits_{-\infty}^{\infty}~dx~\left[ \rho_M(x)+\rho_R(x)    \right]=1~.
\eea
Since, both the densities individually depend explicitly on $p_M$, this condition evaluates $p_M$ exactly. However, $p_M$ can also be computed in the steady state from the following arguments. Intuitively it is easy to get convinced that  $p_M$ is, in fact, the fraction of time the particle spends in phase I in the steady-state. For a detailed discussion on this we refer to the previous observations made in \cite{invariance1,invariance2}. The mean time spent in phase I is $1/r$ (inverse of resetting rate). On the other hand, the time spent while returning to the origin from position $x$ is
nothing but the first passage time to the origin i.e.,
\bea
\tau(x)=[\mathcal{T}=\{\text{inf}: Y(\mathcal{T}|x)=0  \}], 
\eea
where $Y(\mathcal{T}|x)$ depicts the return phase motion which has started from the location $x$ at time zero in its own clock. The return time $\tau(x)$ is a random variable because (i) $x$ is random and determined by the  motion in phase I and (ii) motion of the return phase itself is stochastic namely diffusion in a force field. Thus, we first average over the return phase motion given that it started from $x$ --- this gives us the mean first passage time $\overline{\tau}(x)$. Performing the second average over $x$, weighted by the conditional distribution in phase I, we obtain the mean time spent in the return phase (or equivalently the mean return time)
\bea
\langle \overline{\tau}(x) \rangle= \int\limits_{-\infty}^{\infty}~\mathrm{d}x~ \overline{\tau}(x) \rho(x|\text{\text{phase I}})=\int\limits_{-\infty}^{\infty}~\mathrm{d}x~ \overline{\tau}(x)~\rho_{\text{inst}}(x)~.\hspace{0.3cm}
\label{taux}
\eea
Thus the total average time for a trial of the compound motion is $1/r+\langle \overline{\tau}(x) \rangle$. Hence, the fraction of time spent in phase I motion in steady state is given by
\bea
p_M=\frac{1/r}{1/r+\langle \overline{\tau}(x) \rangle}~.
\label{pm-exact}
\eea
In  \ref{mfpt}, we have reviewed one of the standard methods to compute the mean first passage time $ \overline{\tau}(x)$ and then applied the same to compute the mean return time for two choices of return motions.
\\

\section{Applications to linear and quadratic potentials}
\label{sec:ex}
In this section, we present two exactly solvable models using the general framework presented above. As mentioned earlier, although the formalism described above can be applied to arbitrary Markovian stochastic processes both in phase I and II, we restrict ourselves in this paper to study diffusive processes. More precisely, we consider that in phase I,  the particle  diffuses freely with diffusion constant $D$, thus $\mathcal{L}_0=D\frac{\partial^2}{\partial x^2}$, and in phase II we assume that the particle returns to the origin using diffusion but now with a different diffusion constant $D_R$ in presence of an external confining potential $U(x)$ which applies a force $F(x)=-dU(x)/dx$ on the particle. To compute $\rho_M(x)$, we recall from \eref{equivalence1} that it is proportional to $\rho_{\text{inst}}(x)$, the steady-state density for diffusion with stochastic resetting
and instantaneous returns. This density
 can be obtained by solving \eref{inst-propagator} in the steady state and the resulting solution reads \cite{Restart1,Restart2}
 \bea
 \rho_{\text{inst}}(x)=\frac{\alpha}{2}e^{-\alpha|x|}~,~ \text{with}~~~ \alpha=\sqrt{\frac{r}{D}}~.
 \label{rho-inst}
 \eea
Inserting $\rho_{\text{inst}}(x)$ from above into
\eref{equivalence1} we get 
\bea 
\rho_M(x)=\frac{p_M \alpha}{2}e^{-\alpha|x|}.
\label{rhoD-eqnV}
\eea
We now proceed to compute $\rho_R(x)$ in the presence of $U(x)$.
In this paper, we consider the following two choices for the confining potential: 
(i) linear potential: $U(x)=[\lambda_+\theta(x) +\lambda_- \theta(-x)]~|x|$, where $\lambda_\pm >0$ and (ii) harmonic potential: $U(x)=\frac{1}{2}kx^2$ with $k>0$.  Upon each resetting, the trap is switched on i.e., the potential $U(x)$ is made active and the particle diffuses under the force field $F(x)=-dU(x)/dx$, which is 
(i) $F(x)=\lambda_-\theta(-x)-\lambda_+\theta(x)$ for linear potential and (ii) $F(x)=-kx$ for harmonic potential.

\begin{figure}\centering
\includegraphics[width=4.9cm,height=3.5cm]{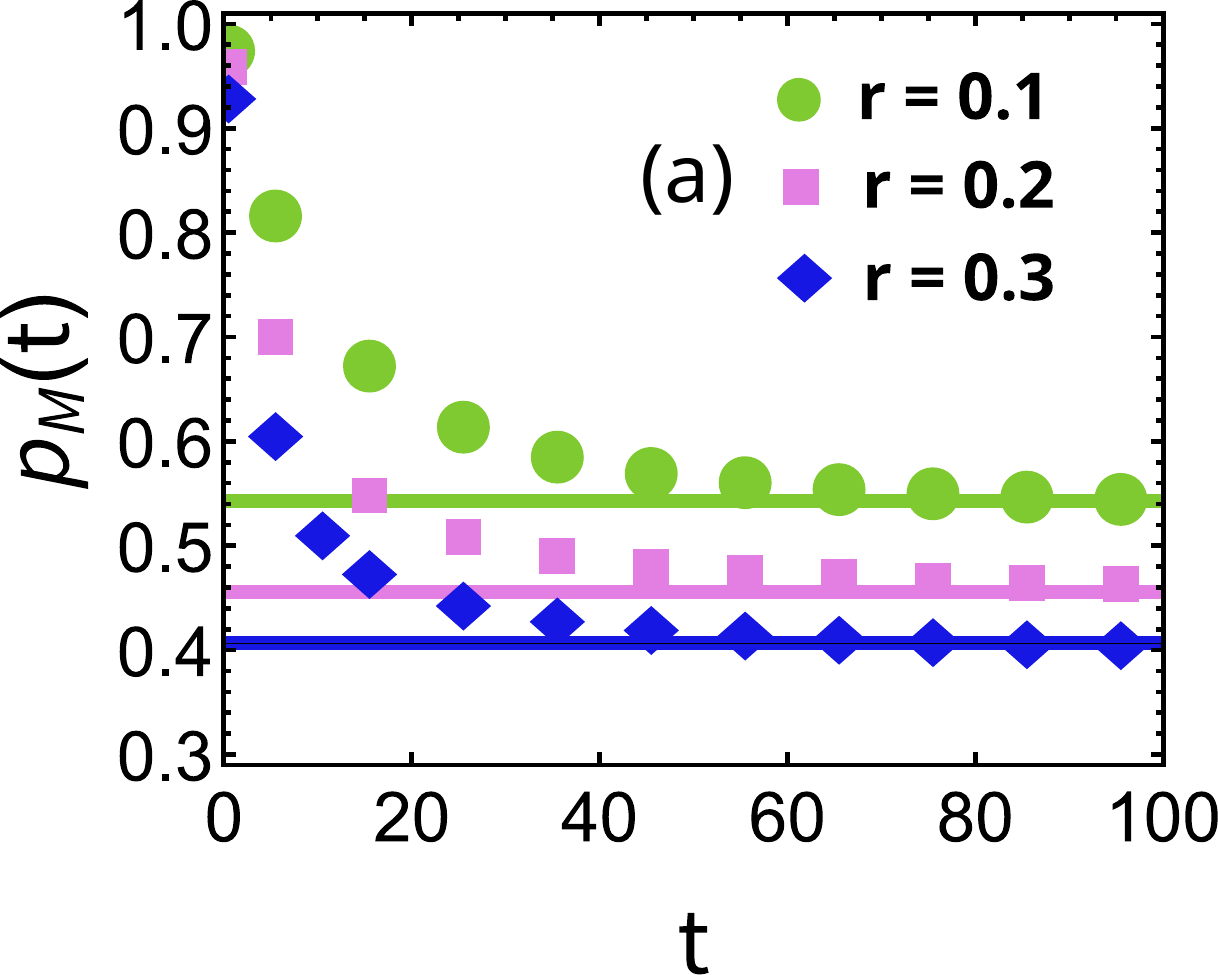}~~~
\includegraphics[width=5.75cm,height=3.5cm]{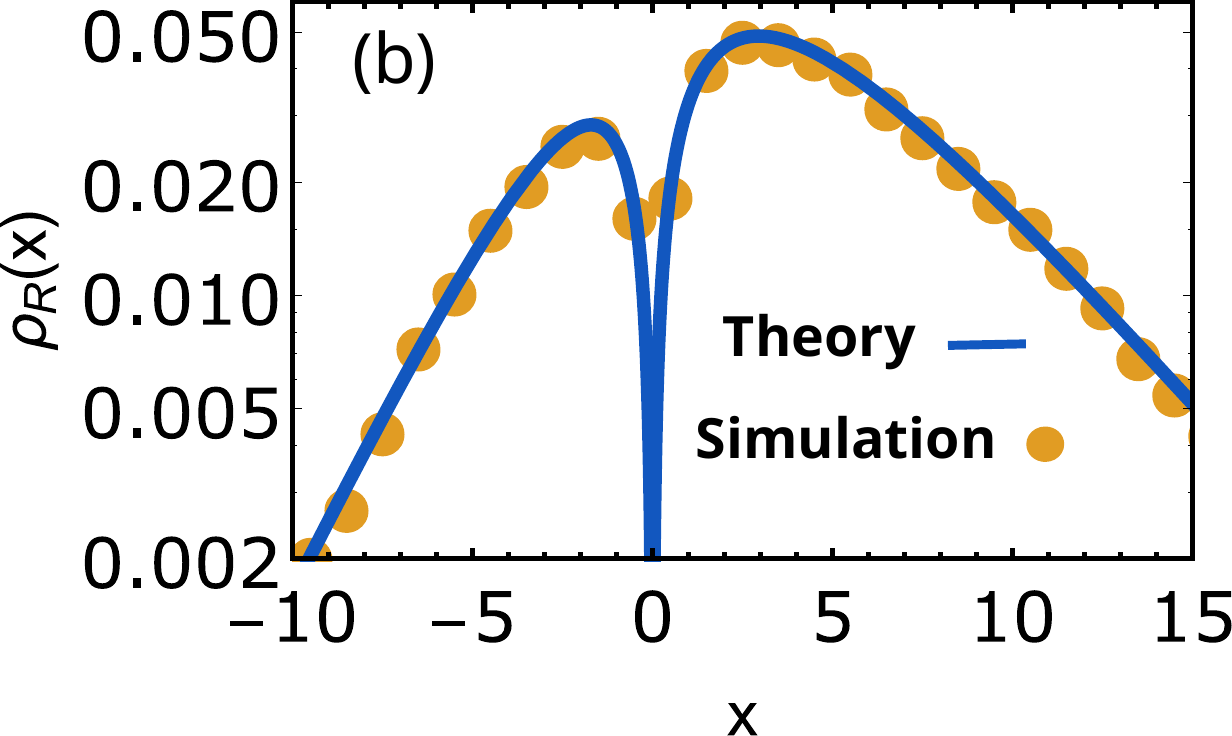}~~~
\includegraphics[width=5.75cm,height=3.5cm]{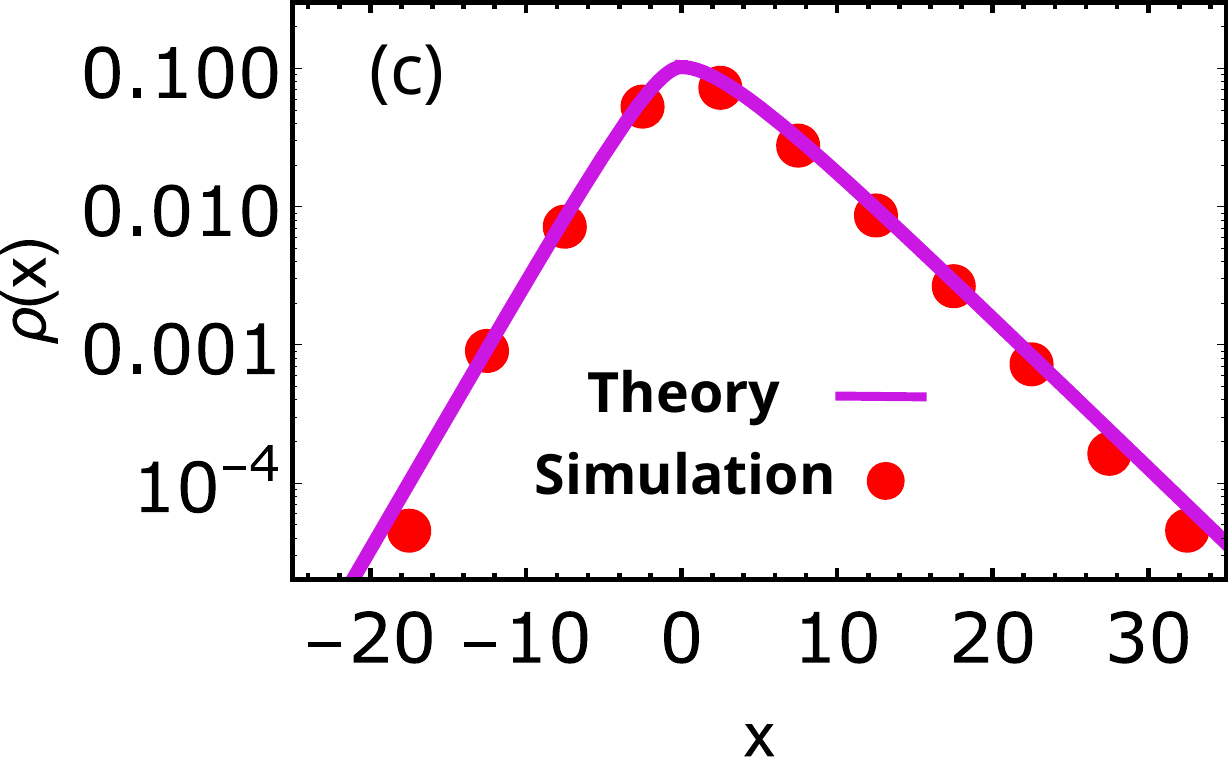}
\caption{Diffusion and stochastic return under the asymmetric force field: $F(x)=\lambda_-\theta(-x)-\lambda_+\theta(x)$. Panel (a): Temporal relaxation of $p_M(t)$ (starting from $p_M(0)=1$) to the steady state, given by \eref{pm-V} and shown by horizontal lines, for three different resetting rates $r=0.1$ (circle), $r=0.2$ (square), and $r=0.3$ (diamond). Panel (b): Comparison of the stationary return density $\rho_R(x)$ given by \eref{eq:sol_R} with numerical simulations.  Panel (c):  Comparison of the total density $\rho(x)$ with the numerical simulations (see \ref{simulation} for the details of the simulation methods). In all plots, simulation data are indicated by symbols while the analytical results are shown by solid lines. For panel (b) and (c), we have set $r=0.2$. Values of other common parameters used in all panels are: $D_R=1$, $D=1$, $\lambda_-=0.75$, and $\lambda_+=0.25$. Numerical simulations were performed for $dt=10^{-3}$ and $10^5$ realizations.}
\label{SS-V}
\end{figure}

\subsection{Linear potential: $U(x)=[\lambda_+\theta(x) +\lambda_- \theta(-x)]~|x|$}
\label{linear_V}
In this case, the particle in phase II experiences a drift $\lambda_+ (\lambda_-)$ towards the origin on positive (negative) side of the origin. To compute the steady state density $\rho_R^\pm(x)$ in the return phase, we now solve \eref{conditional} with the boundary conditions (\eref{rhoR-abs} and \eref{rho-infinity}). This yields the following steady state solutions 
\begin{align}
\begin{split}
\rho_R^-(x)&=\frac{r p_M}{2} \frac{e^{ \alpha x}}{\lambda_--\alpha D_R} \left[ 1-e^{\frac{\lambda_--\alpha D_R}{D_R}x}  \right]~, \\
\rho_R^+(x)&=\frac{r p_M}{2} \frac{e^{- \alpha x}}{\lambda_+-\alpha D_R} \left[ 1-e^{-\frac{\lambda_+-\alpha D_R}{D_R}x}  \right]~.
\end{split}
\label{eq:sol_R}
\end{align}
Combining $\rho_R^-(x)$ and $\rho_R^+(x)$ together as in \eref{jointrho-R} gives us the steady state density $\rho_R(x)$ for the  particles returning to the origin.
In particular, when $\lambda_-=\lambda_+=\lambda$, the full steady state density (using \eref{jointrho-R}) in the return phase reads
\bea
\rho_R(x)=\frac{rp_M}{2} \frac{e^{-\alpha|x|}}{\lambda-\alpha D_R} \left[  1-e^{-\frac{\lambda-\alpha D_R}{D_R}|x|} \right]~.
\label{rhoR-eqnV}
\eea

It is evident that the steady state densities $\rho_M(x)$ and $\rho_R(x)$ in both phases are proportional to the factor $p_M$ which now can be computed from the  normalization of the total density $\rho(x)$, where $\rho(x)=\rho_M(x)+\rho_R^+(x)\theta(x)+\rho_R^-(x)\theta(-x)$. This yields
\bea
p_M=\dfrac{2\alpha \lambda_-\lambda_+}{2\alpha \lambda_-\lambda_++r(\lambda_-+\lambda_+)}~,
\label{pm-V}
\eea
which for symmetric force field becomes
$p_M=\frac{\lambda \alpha}{r+\lambda \alpha}$.

Another way to obtain $p_M$ would be to utilize \eref{pm-exact} which in turn requires the knowledge of the mean return time $\langle \overline{\tau}(x)\rangle$. For a given starting position $x$, the mean first passage time of a diffusing particle to the origin is given by $\overline{\tau}(x)=|x|\left[\frac{\theta(-x)}{\lambda_-}+\frac{\theta(x)}{\lambda_+}\right]$ (see \ref{MFPT-V} for a detailed calculation). Now performing the average of $\overline{\tau}(x)$ over $\rho_{\text{inst}}(x)$, we find the mean return time: $\langle \overline{\tau}(x)\rangle=\frac{1}{2 \alpha}\left(\frac{1}{\lambda_-}+\frac{1}{\lambda_+}\right)$. Substituting $\langle \overline{\tau}(x)\rangle$ into \eref{pm-exact}, we recover the same expression for $p_M$ in \eref{pm-V} as expected. In \fref{SS-V}a, we have shown time dependent behavior of $p_M(t)$ defined in \eref{pa-pb}. As can be seen from the figure, $p_M(t)$ evolves to its steady state value given by \eref{pm-V}. 
Once $p_M$ is known, we have the complete information about the steady state densities. In  
\fref{SS-V}b and \fref{SS-V}c, we observe excellent agreement between the numerical simulation data (see \ref{simulation} for the details of the simulation methods) and the theoretical results for the steady state density $\rho_R(x)$ (given in \eref{eq:sol_R}) and the total density $\rho(x)=\rho_M(x)+\rho_R(x)$, respectively, with $\rho_M(x)$ given by \eref{rhoD-eqnV}.

\begin{figure}\centering
\includegraphics[width=12cm]{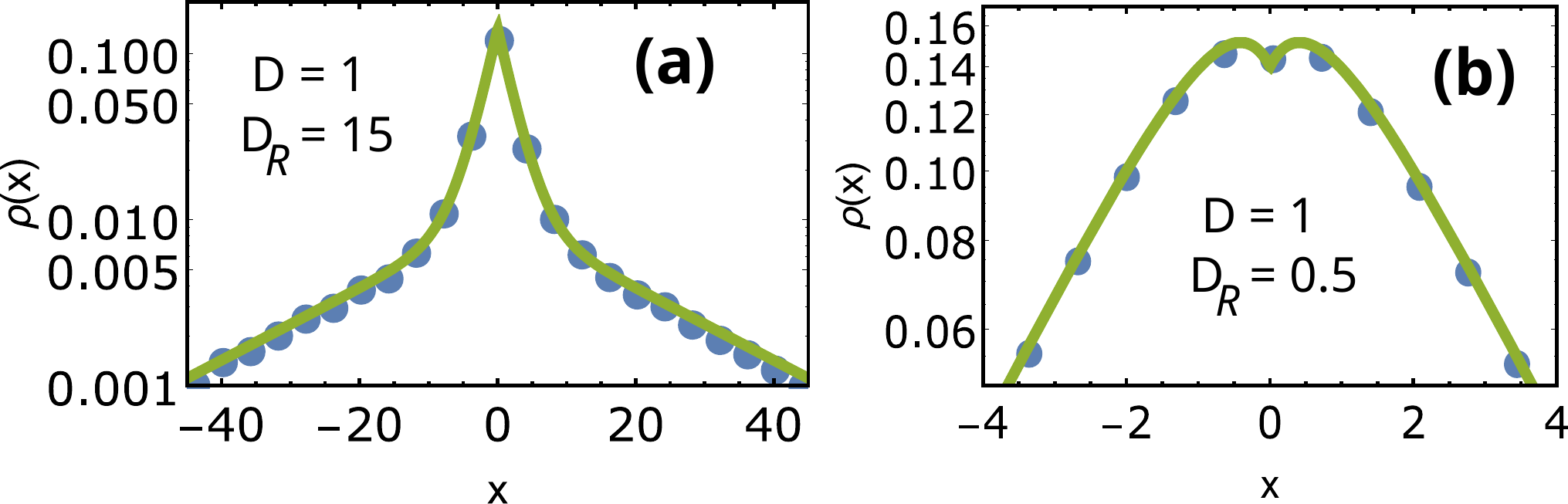}
\caption{Variation in shapes of $\rho(x)$ in the case of returns towards the origin in presence of symmetric linear potential is shown in corroboration with the simulation data when $D\neq D_R$. When $D<D_R$, the slope near the origin is negative (panel a) while for $D>D_R$ the slope near the origin is positive (panel b). In both cases, the density develops exponential tails far away from the origin. Parameters used in these plots are: $\lambda=0.75, r=0.2$. Numerical simulations were performed for $dt=10^{-3}$ and $10^6$ realizations.}
\label{SS-V-limits}
\end{figure}

Several comments are in order now. The total density exhibits interesting asymptotic behaviors in the limit of small and large $x$. In the following, we furthermore confine our analysis to the symmetric force case i.e., when $\lambda_\pm=\lambda$ (however the analysis can be easily generalised to the asymmetric cases). We first discuss the behaviour near $x=0$. Expanding $\rho_R(x)$ and $\rho_M(x)$ in \eref{rhoR-eqnV} and \eref{rhoD-eqnV}, respectively, around $x=0$, we get 
$\rho_R(x) \approx \frac{rp_M}{2D_R}|x|-\frac{rp_M(\lambda+\alpha D_R)}{4D_R^2}x^2$ and 
$\rho_M(x)\approx \frac{\alpha p_M}{2}-\frac{\alpha^2 p_M}{2}|x|+\frac{\alpha^3 p_M}{4}x^2$ for small $x$. Plugging into the 
total density we get 
\begin{align}
\rho(x) \approx \frac{\alpha p_M}{2} \left[ 1+\alpha \left(\frac{D}{D_R}-1  \right)|x|  \right] 
+ \frac{\alpha^2 p_M x^2}{4} \left[ \alpha \left( 1-\frac{D}{D_R} \right) -\frac{\lambda D}{D_R^2}\right]~,
\label{rho-limit-small-x}
\end{align}
for small $x$.
Note that when $D=D_R$, the linear term in $\rho(x)$ does not contribute and the profile is parabolic near the origin as shown in \fref{SS-V}c. On the other hand, when $D \neq D_R$, the linear term survives developing a cusp at the origin. See \fref{SS-V-limits} where we have plotted the steady state for $D<D_R$ (panel a) and $D>D_R$ (panel b). The cusp at the origin is clearly visible and the slope near the origin then depends on the ratio $D/D_R$. 

  The large $x$-behavior of the densities depends on both $\lambda$ and $\alpha D_R$. For $\lambda/D_R>\alpha$, the total density behaves as
\bea
\rho(x) \approx \frac{\alpha p_M}{2} \left[ 1+\frac{\alpha D}{\lambda -\alpha D_R}  \right] e^{-\alpha |x|} ~.
\eea
Thus the tail is controlled by the length scale $\alpha^{-1}$, which is the typical (also average) distance traversed by the particle in phase I.
On the other hand in the limit $\lambda/D_R<\alpha$, the total density behaves as
\bea
\rho(x) \approx \frac{\alpha p_M}{2} \left[ \frac{\alpha D}{\alpha D_R-\lambda}  \right] e^{-\frac{\lambda}{D_R}  |x|}~,
\eea
where the tail is effected by the factor $\lambda/D_R$. This exponential behaviour is verified numerically in \fref{SS-V-limits}a.

In summary, we find that near the origin, the total density has either linear or quadratic behavior while in the large $x$, the densities have exponential tails.  The analysis above holds when $\lambda \neq \alpha D_R$. However, the singular limit $\lambda \to \alpha D_R$ has to be treated carefully.  In this limit, we find that the return density from \eref{rhoR-eqnV} simplifies to $\rho_R(x) = \frac{rp_M~}{2 D_R}  |x| e^{-\alpha |x|}$. Nevertheless, the asymptotic behaviour of $\rho(x)$ for small $x$ and large $x$ remains same as before.

\begin{figure}\centering
\includegraphics[width=4.9cm,height=3.5cm]{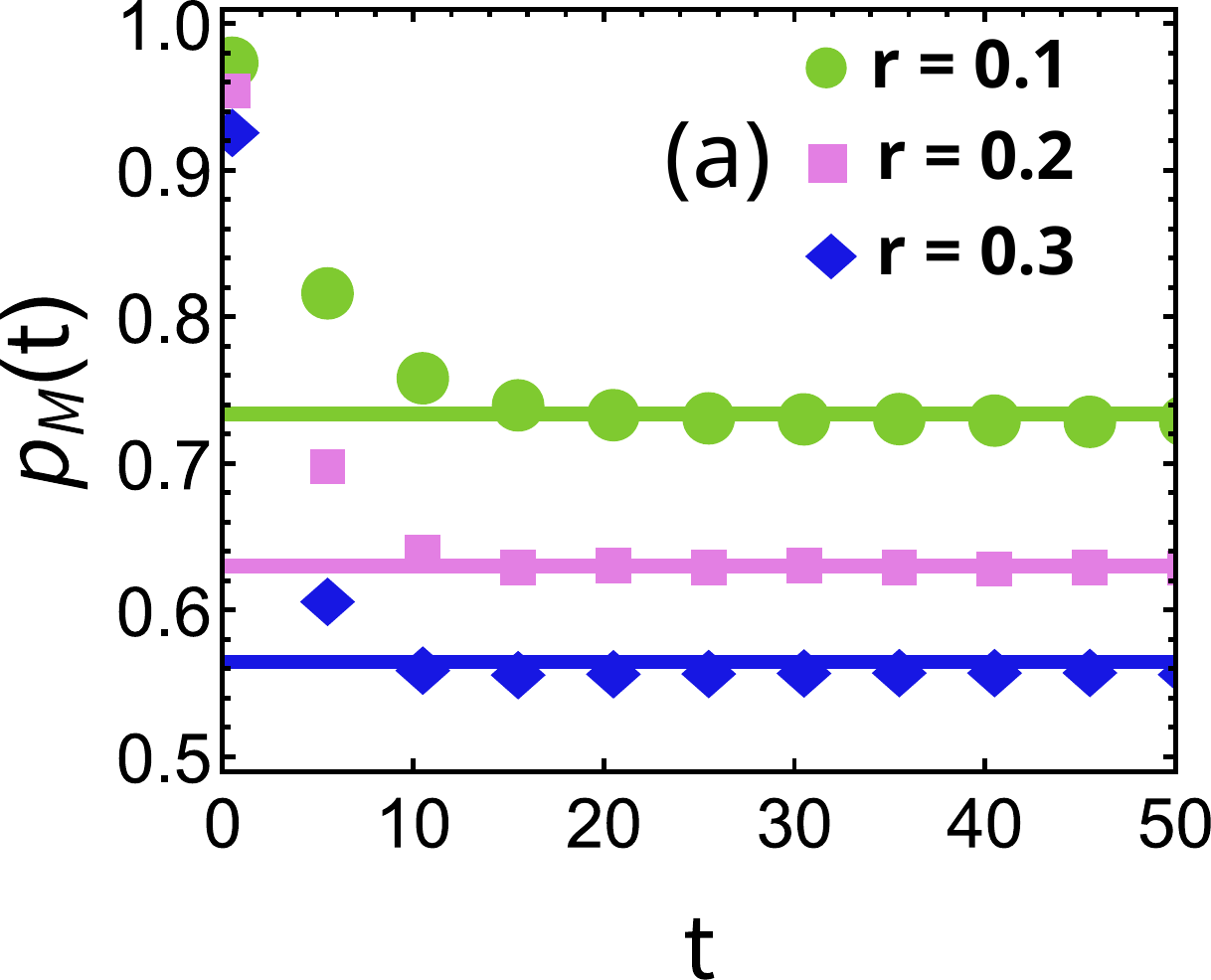}~~~
\includegraphics[width=5.75cm,height=3.5cm]{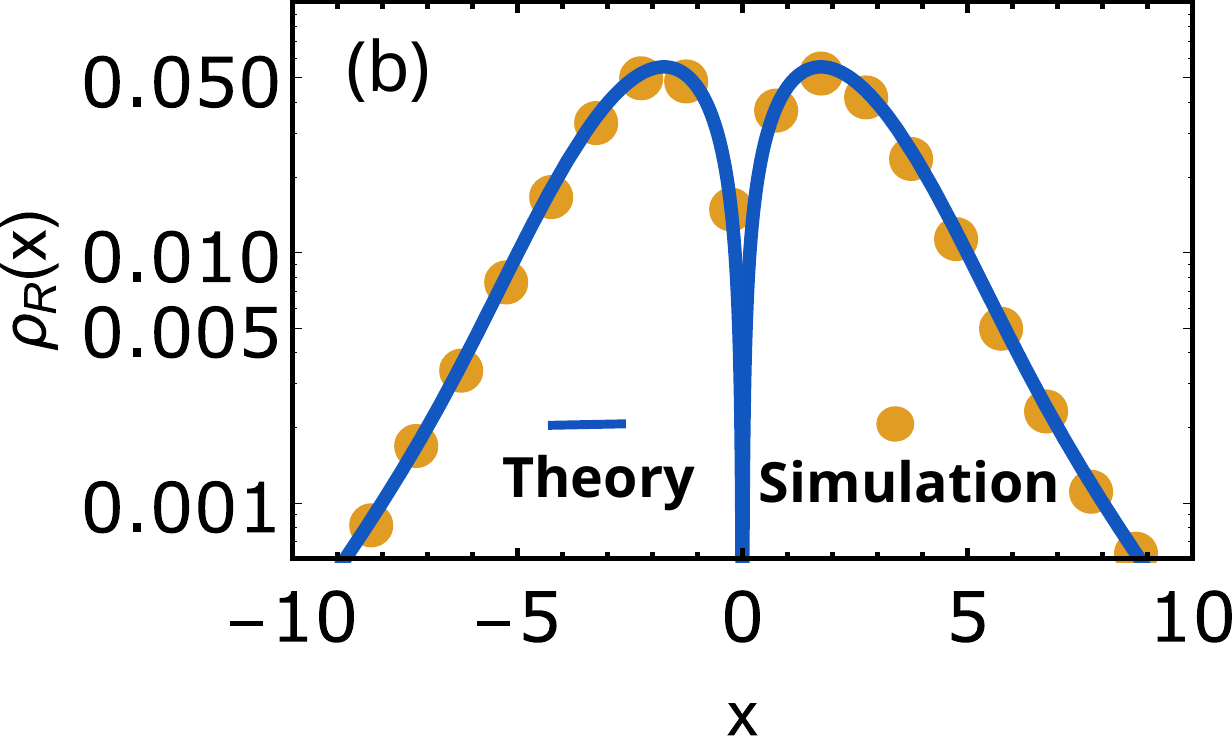}~~~
\includegraphics[width=5.7cm,height=3.5cm]{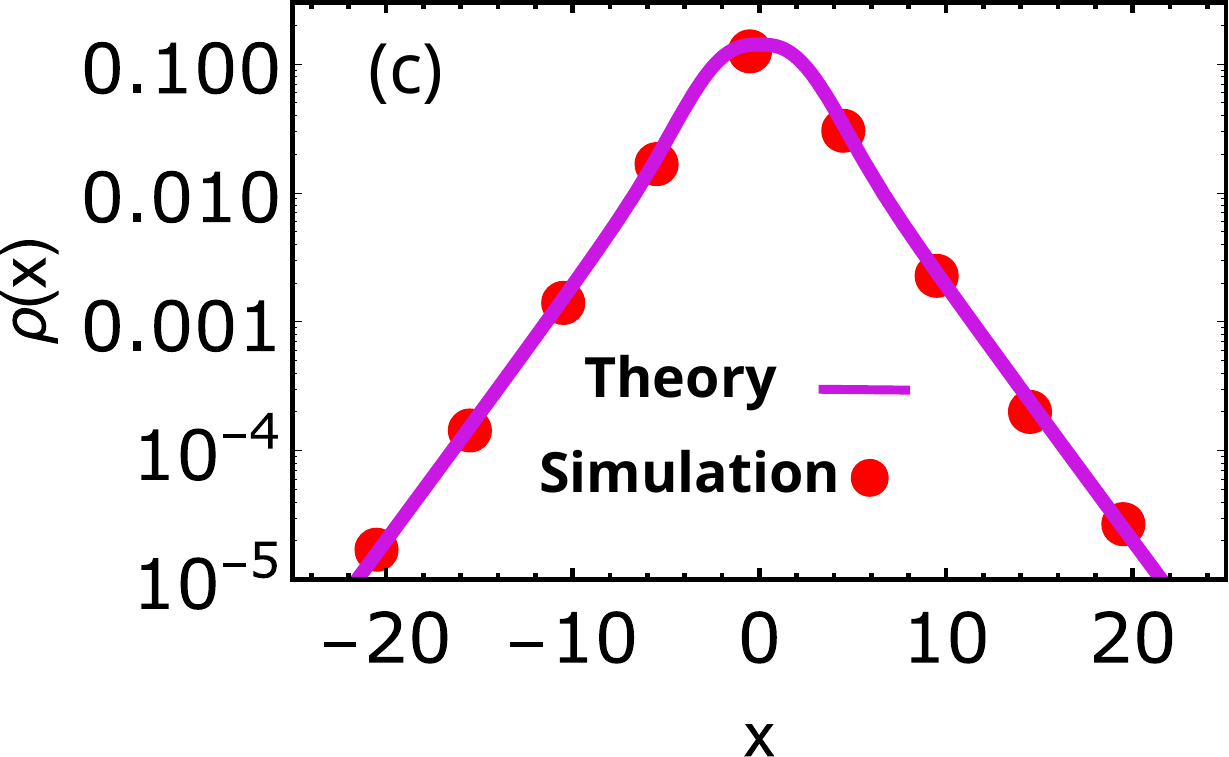}
\caption{Diffusion and stochastic return under the force field $F(x)=-kx$. Panel (a): Temporal relaxation of $p_M(t)$ (starting from $p_M(0)=1$) to the steady state values (computed from the theory and shown by solid horizontal lines) for three different resetting rates $r=0.1$ (circle), $r=0.2$ (square), and $r=0.3$ (diamond). Panel (b) and panel (c) respectively compare theoretical expressions for the steady state return density $\rho_R(x)$ (given by \eqref{rhoR-H1}) and total density $\rho(x)$ with the simulation data (see \ref{simulation} for the details of the simulation methods). In all plots, simulation points are denoted by symbols while the theory results are indicated by solid lines. We have set $r=0.2$ in panel (b) and panel (c). In all panels, we have used the following set of parameters: $D_R=1$, $D=1$, and $k=0.3$.  Numerical simulations were done for $dt=10^{-3}$, and $10^5$ (panel a), $10^6$ (panel b \& panel c)  realizations respectively.}
\label{SS-H}
\end{figure}

\subsection{Harmonic potential: $U(x)=k x^2/2$}
\label{sec:har-pot}
We now consider the case when the return process is driven by the harmonic potential $U(x)=\frac{1}{2}kx^2$. Such harmonic potential is possible to realize in experiments using the standard optical trap technique \cite{optical-trap}.
As mentioned before, in this case also the steady density for the particles diffusing freely in phase I is given by \eref{rhoD-eqnV}. On the other hand, governing equations for the returning particles follow from \eref{conditional}
\bea
-\partial_x \left[ J_R^{\pm}(x)  \right]+r\rho_M(x)=0~,
\label{rhoM-H1}
\eea
where
\bea
J_R^{\pm}(x)=-kx \rho^{\pm}_R(x)-D_R \frac{\partial  \rho^{\pm}_R(x)}{\partial x}~,
\label{currentM-H1}
\eea
are the probability currents in the return phase. We present the solutions for $\rho_R(x)$ in the positive $x$-axis. The solutions in the negative $x$-axis will be identical due to the underlying symmetry of the problem. We start by integrating \eref{rhoM-H1} along with \eref{currentM-H1} on the positive $x$-axis. This gives
\bea
J_R^+(x)=c_2-\frac{rp_M}{2}e^{-\alpha x}~,
\eea
where $c_2$ is a constant and is found to be zero by setting the boundary condition $J_R^+(x\to \infty)=0$ (see \sref{bcss}). To solve the resulting equation for the current $J_R^+(x)=-\frac{rp_M}{2D_R}e^{-\alpha x}$, we explicitly use \eref{currentM-H1} and find
\bea
\rho_R^+(x)=c_1~e^{\frac{-kx^2}{2D_R}}-\frac{rp_M}{2}\sqrt{\frac{\pi}{2kD_R}}e^{\frac{-D_R \alpha^2}{2k}} e^{\frac{-kx^2}{2D_R}} \text{Erfi}\left[ \frac{-kx+\alpha D_R}{\sqrt{2kD_R}} \right].~ \hspace{0.1cm}
\eea
To evaluate the integration constant $c_1$, we now use the boundary condition \eref{rhoR-abs} (see  \sref{bcss}). A similar solution can also be obtained in the negative real axis. Combining together, we finally arrive at the steady state density for the particles in return phase
\begin{align}
\rho_R(x)=\frac{rp_M}{2}\sqrt{\frac{\pi}{2kD_R}}e^{\frac{-D_R \alpha^2}{2k}} e^{\frac{-kx^2}{2D_R}} \bigg( \text{Erfi}\left[\alpha\sqrt{\frac{D_R}{2k}} \right]- \text{Erfi}\left[ \frac{-k|x|+\alpha D_R}{\sqrt{2kD_R}} \right] \bigg).~
\label{rhoR-H1}
\end{align}
Equations \eref{rhoD-eqnV} and \eref{rhoR-H1} constitute the full steady state solutions of a diffusing particle under returns modulated by the harmonic potential. The only unknown constant $p_M$, can again be obtained by invoking the normalization condition of the total density $\rho(x)$ as before. However, it is not possible to get a closed form expression for $p_M$ like the linear potential case as shown in the previous section. Hence, we proceed to solve for $p_M$ numerically. 

For completeness, we also discuss the alternative approach to compute $p_M$ as put forward in \sref{pm}. Recall that $p_M$ is the fraction of time spent by the particle in the diffusive phase in the steady state. Thus one requires to find the mean first passage time $\overline{\tau}(x)$ of a particle in a harmonic potential to the origin starting from the position $x$. This is rather a lengthy calculation, thus we leave the details in the appendix and just present the final result here. Following \ref{MFPT-H}, we find
\bea
\overline{\tau}(x)=\frac{ (\gamma+\ln{4}) - \mathcal{U}^{(1,0,0)}\left(0,\frac{1}{2},\frac{\gamma_R  x^2}{2}\right) }{2D_R \gamma_R }>0~, \hspace{0.3cm}
\label{tau-H}
\eea
where $\gamma_R=k/D_R$, $\gamma$ is the Euler's constant and $\mathcal{U}^{(1,0,0)}(a,b,z)$ is the partial derivative of the Tricomi confluent hypergeometric function $\mathcal{U}(a,b,z)$ with respect to $a$ \cite{ref-1,ref-2}.
Due to the underlying homogeneity of the problem, note that $\overline{\tau}(x)$ is symmetric.
Averaging $\overline{\tau}(x)$ over the steady state density $\rho_{\text{inst}}(x)$, we finally obtain $\langle  \overline{\tau}(x) \rangle$. Note that this integration cannot be carried out analytically, hence we integrate numerically. We then plug $\langle  \overline{\tau}(x) \rangle$ into the expression for $p_M$ (given by \eref{pm-exact}) in the steady state and verified with the numerical simulations. In \fref{SS-H}a, we have demonstrated relaxation of $p_M(t)$ as a function of time to its steady state value for three different resetting rates. \fref{SS-H}b and \fref{SS-H}c respectively show agreement of analytical results for the steady state return density $\rho_R(x)$ in the return phase and the total density $\rho(x)$ with the numerical simulations. 

\begin{figure}\centering
\includegraphics[width=12cm]{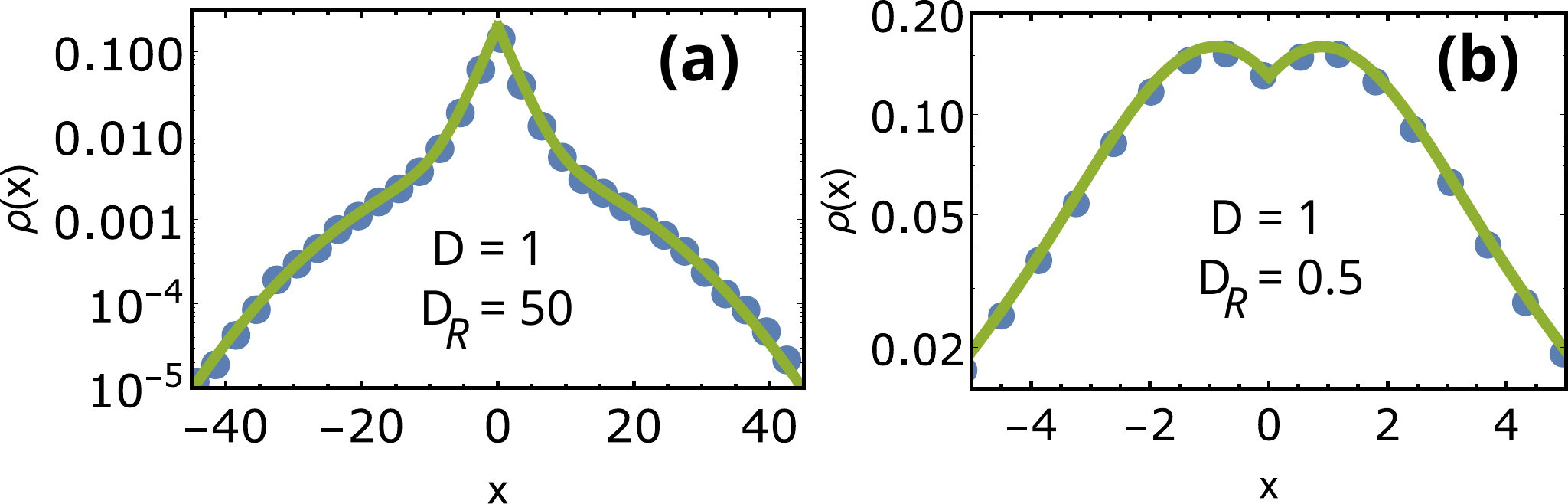}
\caption{Variation in shapes of $\rho(x)$ for diffusion with stochastic returns modulated by harmonic potential is shown in comparison with the simulation data. We have considered two scenarios: (i) $D<D_R$ (panel a) -- the slope near the origin is negative (see \eref{rhoHsmallxlimit}) and (ii) $D<D_R$ (panel b) -- the slope near the origin is positive (see \eref{rhoHsmallxlimit}). Panel (a) shows a \textcolor{blue}{deviation} from the linear behavior while the exponential tail appears at far away from the origin where the probability is small and hard to access by normal Monte-Carlo simulations. On the other hand, in panel (b), we see that $\rho(x)$ develops an exponential tail away from the origin. Parameters used here are: $k=0.3,~r=0.2$. Numerical simulations are performed for $dt=10^{-3}$ and $10^6$ realizations.}
\label{SS-H-limits}
\end{figure}

We now turn our attention to discuss the asymptotic behaviors of the steady state densities in the limits of small and large $x$ where again
interesting observations can be made. In the small $x$ limit, expanding the total density around $x=0$, we get
\bea
\rho(x) &\approx& 
\frac{\alpha p_M}{2} \left[ 1+\alpha \left( \frac{D}{D_R} -1\right)|x|  \right]+\frac{\alpha^3 p_M}{4}\left[ 1- \frac{D}{D_R}\right]  x^2 \nonumber \\
&+& \frac{\alpha^2 p_M}{12 D_R^2} \left[ \alpha^2 D D_R-\alpha^2 D_R^2-2k D \right]x^2|x|~,
\label{rhoHsmallxlimit}
\eea
where we have used the following small $|x|$ asymptotic behaviour
\bea
\text{Erfi}\left[ \frac{a-b|x|}{c} \right] \approx \text{Erfi}\left[\frac{a}{c} \right]-\frac{2 b |x|}{c \sqrt{\pi}}~ e^{a^2/c^2}~.
\eea
Note that when $D=D_R$, both the linear and second order terms in \eref{rhoHsmallxlimit} drop out, thus the first and second order derivatives vanish at the origin. The leading contribution in this case comes from the third order term (see \fref{SS-H}c). However, when $D_R \neq D$, we see the leading behavior (i.e., the slope) near the origin is determined by the linear term, and consequently one sees a deviation from the linear behavior (see \fref{SS-H-limits}). Various interesting shapes of the total density $\rho(x)$ are presented in \fref{SS-H-limits}.

\begin{figure}\centering
\includegraphics[width=9cm]{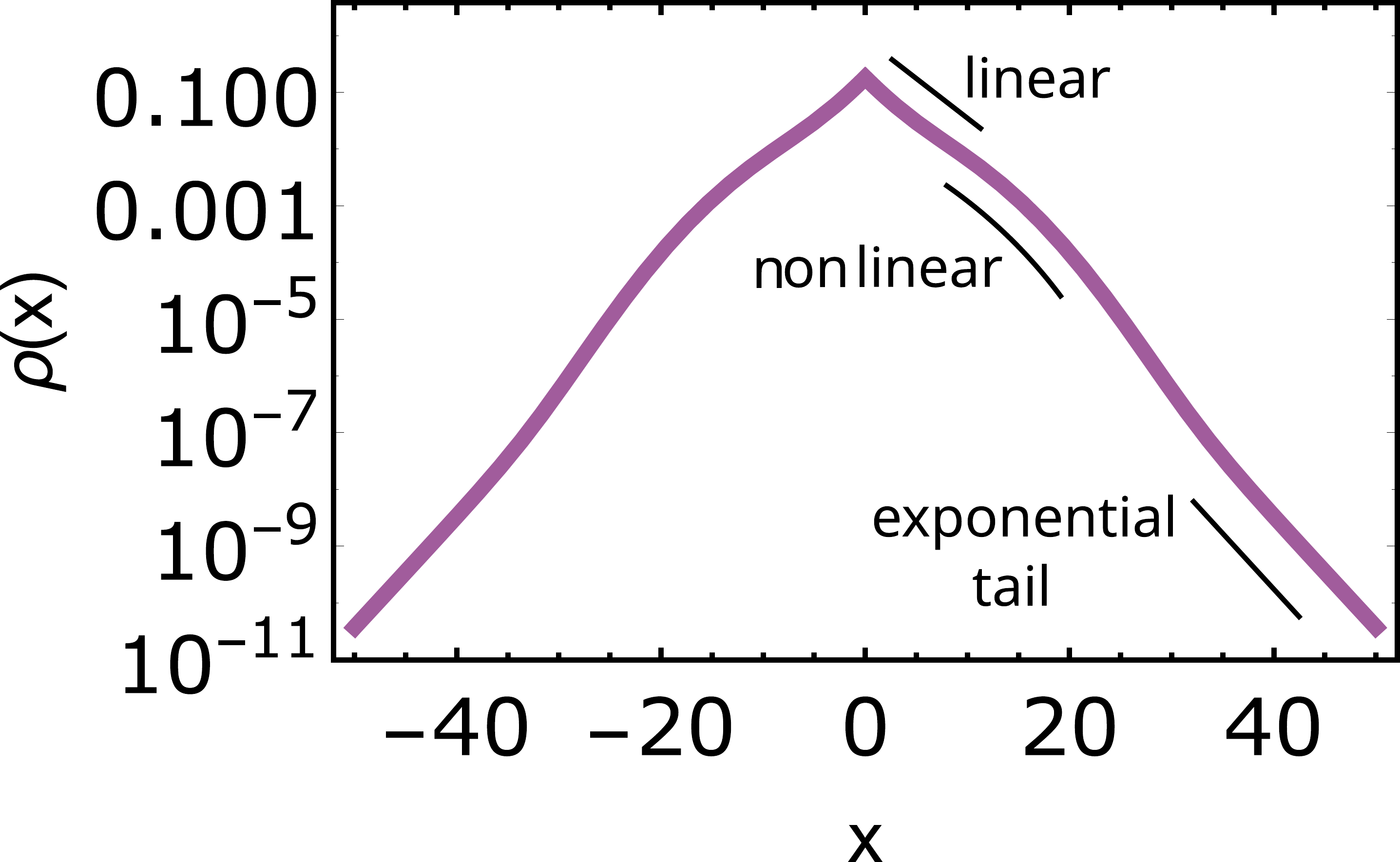}
\caption{Behavior of the steady state density in the case of returns driven by harmonic potential $U(x)=\frac{1}{2}kx^2$ when $D\neq D_R$. We observe three different regimes in the steady state: linear (at small $x$), non-linear (intermediate) and an exponential tail (at large $x$). Parameters: $D=1,~D_R=12,~k=0.3,~r=0.2$.}
\label{SS-H-all-behavior}
\end{figure}

We now proceed to perform the asymptotic analysis for large $x$. We first note that $\rho_R(x)$ in \eref{rhoR-H1} can be written as a difference between two terms i.e., $\rho_R(x)=\rho_R^{I}(x)-\rho_R^{II}(x)$, where $\rho_R^{I(II)}(x)$ is the first (second) term on the RHS of \eref{rhoR-H1}. We find that for large $x$,  $\rho_R^{II}(x)$ behaves asymptotically as 
\bea
\rho_R^{II}(x) &\approx& \frac{rp_M}{2} \sqrt{\frac{\pi} {2 \pi k D_R}} e^{-\frac{D_R \alpha^2}{2k}}  e^{-\frac{kx^2}{2D_R}} \frac{\sqrt{2kD_R}}{\alpha D_R-k|x|} \exp \left[ \left( \frac{-k|x|+\alpha D_R}{\sqrt{2k D_R}} \right)^2 \right]\nonumber \\
&\approx& \frac{\alpha p_M}{2} e^{-\alpha|x|} \frac{\alpha D}{\alpha D_R-k|x|}~.
\eea
Here note that $\rho_R^I(x)$ decays faster $(\sim e^{-x^2})$ than $\rho_R^{II}(x)$, hence can be neglected altogether. Combining all the individual contributions together, we find
\bea
\rho(x) \approx \frac{\alpha p_M}{2} e^{-\alpha|x|}, 
\eea
which again has an exponential tail in the large $|x|$ limit. We refer to \fref{SS-H-all-behavior} which shows a typical steady state distribution in this case exhibiting the linear behaviour at small $|x|$, then a crossover to an intermediate nonlinear behaviour before developing the exponential tail at large $|x|$.

\subsection{$D_R \to 0$ limit}
In this subsection, we discuss the $D_R \to 0$ limit which, in our case, makes the return protocols deterministic. In the first example of linear potential, this would mean that the particle is being pulled at a constant drift towards the resetting location. To see this, we take the limit in \eref{rhoR-eqnV} to find $\rho_R(x) \approx \frac{rp_M}{2\lambda}e^{-\alpha|x|}$. Note that this result was already reported in \cite{invariance2} which was obtained by solving \eref{conditional} with $D_R=0$. For the harmonic case, by taking $D_R \to 0$ limit in \eref{rhoR-H1}, we find   $\rho_R(x) \approx \frac{rp_M}{2k|x|}e^{-\alpha|x|}$ where we have used the following asymptotic forms of the Erfi-function namely $\text{Erfi}[x \to 0] \approx \frac{2x}{\sqrt{\pi}}$ and $\text{Erfi}\left[ |x| \to \infty \right] \approx \frac{1}{\sqrt{\pi}} \frac{e^{x^2}}{|x|}$. Again, this result can also be obtained by first putting  $D_R=0$ in \eref{conditional} and then solving $\frac{\partial}{\partial x}[kx \rho_R(x)]+ \frac{r\alpha p_M}{2} e^{-\alpha |x|}=0$ with natural boundary conditions $\lim_{x \to  \pm \infty} \rho_R(x)=0$. Note that the return density solutions obtained in this way take non-zero values when $x=0$. This is in contrast with the vanishing values obtained when evaluating the solutions (e.g., \eref{rhoR-eqnV} and \eref{rhoR-H1}) for $D_R\neq 0$, which is consistent with the boundary conditions  in \eref{rhoR-abs}. Thus, we infer that the limits $x \to 0$ and $D_R \to 0$ do not commute and  
it may be related to the fact that the governing master equations in return phase (see e.g., \eref{rho-R-full}) change order in the sense that they become first order differential equations as derived for the deterministic return protocols \cite{invariance1,invariance2}.


\section{Conclusions}
\label{sec:concl}
In this paper, we have provided a formulation to study stochastic resetting processes which are subject to non-instantaneous returns. We have investigated the possibility of stochastic return protocols which were so far unexplored. We concur that in physical world microscopic processes are sensitive to thermal fluctuations and thus the return processes can also be random despite measurements with extreme precision. We have developed a comprehensive framework for such a set up which then systematically allows us to derive the governing equations for the observables like position herein, and potentially others in general.

In particular, our framework allows one to study the time evolution of the full process which can be decomposed into two stochastic phases. For each phase, we have derived Fokker Planck equations which are coupled through the probability currents to alternate between the phases. Utilizing our formalism, we then demonstrate how to capture the novel aspects of the position distributions especially in the steady state limit. Quite interestingly, we observed that some of the parameters emerged from the solutions can be interpreted in terms of the mean residence time in each phase.

It is important to emphasize that our approach is not only limited to diffusion in the  phase I, but can easily be extended to a wide range of stochastic motions beyond diffusion (generated from an arbitrary Fokker-Plank operator $\mathcal{L}_0$). Furthermore, our formalism can readily be applied to settings with more general trapping potentials $U(x)$ other than linear or harmonic potential in phase II (see \eref{rhoR-eqn2} and \eref{rhoD-eqn}). Thus, we believe that the framework put forward here offers a widespread applicability for the theory as well as future experimental studies of stochastic resetting.

Various generalizations of this problem can be immediately invited. While the thermodynamics of instantaneous resetting is only getting started \cite{thermo1,thermo2,thermo3,thermo4,thermo5}, a future direction would be also to study thermodynamics of the  models studied above. From the dynamical perspective, extreme statistics of resetting process has hardly been explored. It therefore remains to be seen how the extremal properties e.g., the statistics of largest excursion \cite{extreme} will be ramified in the presence of resetting. Another interesting aspect would be to study the problem in the presence of an absorbing boundary or a target at a distance away from the resetting location. One could investigate the first passage properties such as the survival probability or the mean first passage time to the target in the presence of stochastic returns.  Advantage can be taken by noting that the whole compound process gets renewed after each trial, and thus a renewal formalism applied to the full process seems to be a promising direction. In fact, if we assume that target detection is possible by the searcher only during motion phase, but not in return phase, one could apply the renewal structure as in stochastic resetting systems with refractory periods \cite{HRS,refractory} to obtain the first passage time statistics.

Remarkably, implementation of our resetting protocol does not require to track the particle during its return. A counter could be set which can detect when the particle makes a first passage to the origin. We refer to another recent study \cite{intermittent-potential} that proposes similar ideas of diffusion with stochastic returns using intermittent resetting potentials.

\section{Acknowledgements}
D. G. is supported by “Excellence Project 2018” of the Cariparo foundation. C. A. P. acknowledges the support from University of Padova through Project No. STARS-Stg (CdA Rep. 40, 23.02.2018) BioReACT grant. A.K. acknowledges support from the SERB Early Career Research
Award ECR/2017/000634 from the Science and Engineering Research Board,
Department of Science and Technology and the support of the Department of Atomic Energy, Government of India, under project no. 12-R \& D-TFR-5.10-1100. A.P. gratefully acknowledges Raymond and Beverly Sackler postdoctoral fellowship and Ratner Center for Single Molecule Science at the Tel Aviv University for funding.


\appendix

\section{ Mean first passage times of a Brownian particle in linear and quadratic potential}
\label{mfpt}
In this section, we present results for the mean first passage time in the context of our current problems.
There are many standard ways to compute mean first passage time (see \cite{RednerBook,Schehr-review,Redner-review,Grebenkov,Benichou} for review and applications of the subject). Here, we adapt the approach of
backward Fokker Planck equation which is often advantageous to treat the first passage properties \cite{RednerBook,Schehr-review}. Within this context, we ask a simple question: given that a particle started its dynamics from the position $x$ at time zero, how much time $\tau(x)$ it will take to reach some threshold (in our case, the origin) for the first time. To compute the statistical properties of $\tau(x)$ it is convenient to consider the cumulative probability 
$Q(x,t)=\text{Prob.}[\tau(x)\geq t]$. Note that the probability $Q(x,t)$ is actually the survival probability that the particle, starting from $x$, survives from an absorbing boundary at the origin till time $t$. In other words, this is the probability that the particle does not hit the origin till time $t$. The first passage time density, denoted by $f_{\text{FP}}(\tau)$, is then given by $f_{\text{FP}}(\tau)= -\frac{\partial Q(x,\tau)}{\partial \tau}$. Hence the mean first passage time to the origin is given by 
\bea
\overline{\tau}(x)=\int_0^\infty~dt~t\left[- \frac{\partial Q(x,t)}{\partial t} \right].
\label{MFPT-Main0}
\eea

In the backward Fokker-Planck approach, one first considers the initial position as a variable, and then solves the backward equation with suitable boundary conditions self-consistently. The governing equation for the survival probability is given by \cite{RednerBook,Schehr-review}
\bea
\frac{\partial Q(x,t)}{\partial t}=F(x)~\frac{\partial Q(x,t)}{\partial x}+D_R\frac{\partial^2 Q(x,t)}{\partial x^2},
\label{BFP-Q}
\eea
with the initial condition $Q(x,0)=1$ and
the boundary conditions $Q(0,t)=0$. The Laplace transform $\tilde{Q}(x,s)=\int_0^\infty ~dt~ e^{-st}Q(x,t)$ of the survival probability then satisfies the following equation
\bea
-1+s\tilde{Q}(x,s)=D_R\frac{\partial^2 \tilde{Q}(x,s)}{\partial x^2}+F(x) \frac{\partial \tilde{Q}(x,s)}{\partial x}.
\label{BFP-Q-LT}
\eea
It is easy to see that the boundary conditions mentioned above now gets modified to $\tilde{Q}(0,s)=0$ and moreover we used $\lim_{x \to \pm \infty} \tilde{Q}(x,s)< \infty$. Note from \eref{MFPT-Main0} that the mean first passage time can simply be obtained from the $s \to 0$ limit i.e.,
\bea
\overline{\tau}(x) =\tilde{Q}(x,s \to 0),
\label{MFPT-Main}
\eea
where the overline simply represents the average over the stochastic trajectories which start from a fixed position $x$ and we have assumed $\lim_{t \to \infty} tQ(x,t)=0$. However, in our problem we are required to perform an additional average over $x$, which we denote by $\langle \overline{\tau}(x)\rangle$. As discussed in the main text, this average is done over $\rho_{\text{inst}}(x)$. 

Having sketched the basic steps, we are now ready to discuss the following two cases: (i) linear potential $U(x)=[\lambda_+\theta(x) +\lambda_- \theta(-x)]~|x|$, where $\lambda_\pm >0$ and (ii) harmonic potential $U(x)=\frac{1}{2}kx^2$ with $k>0$, as done in the main text.

\subsection{$F(x)=\lambda_-\theta(-x)-\lambda_+\theta(x)$} \label{MFPT-V}
We first consider linear potential case in which a Brownian particle which diffuses in a force field $F(x)=\lambda_-\theta(-x)-\lambda_+\theta(x)$. We want to compute the mean first passage time of this particle to the origin. Applying \eref{BFP-Q-LT}, we find
\bea
D_R\frac{\partial^2 \tilde{Q}(x,s)}{\partial x^2}+\left[ \lambda_-\theta(-x)-\lambda_+\theta(x)  \right]\frac{\partial \tilde{Q}(x,s)}{\partial x}=-1+s\tilde{Q}(x,s)~. \nonumber
\eea
Solving the above equations with the boundary conditions, we find
\bea
\tilde{Q}(x,s)&=&~\frac{1}{s}\left[ 1-e^{-\frac{\sqrt{\lambda_+^2+4D_Rs}-\lambda_+}{2D_R}x}  \right],~x>0~, \nonumber \\
\tilde{Q}(x,s)&=&~~\frac{1}{s}\left[ 1-e^{\frac{\sqrt{\lambda_-^2+4D_Rs}-\lambda_-}{2D_R}x}  \right],~x<0~.
\label{Q-V}
\eea
Finally, utilizing \eref{MFPT-Main} and substituting expressions for $\tilde{Q}(x,s)$ from \eref{Q-V} we arrive at the following expression for the mean first passage time
\bea
\overline{\tau}(x)=|x| \left[ \frac{\theta(x)}{\lambda_+}+\frac{\theta(-x)}{\lambda_-}  \right]~,
\eea
which was used in \sref{linear_V} to compute the mean return time. Finally substituting the above expression in \eref{taux} with the use of \eref{rho-inst}, we get
\bea
\langle \overline{\tau}(x) \rangle=\int \limits_{-\infty}^\infty~dx~\overline{\tau}(x) \left[  \frac{\alpha}{2}e^{-\alpha |x|}  \right]=\frac{1}{2\alpha} \left( \frac{1}{\lambda_+}+\frac{1}{\lambda_-}  \right)~,
\eea
which was announced in the main text (see \sref{linear_V}).

\subsection{$F(x)=-kx$}
\label{MFPT-H}
In this section, we move on to compute the mean first passage time of a Brownian particle diffusing in a potential $U(x)=\frac{1}{2}kx^2$.
We again start with the Laplace space backward Fokker-Planck equation for the survival probability
\bea
D_R\frac{\partial^2 \tilde{Q}(x,s)}{\partial x^2}-kx \frac{\partial \tilde{Q}(x,s)}{\partial x}=-1+s\tilde{Q}(x,s)~,
\eea
with the boundary conditions $\tilde{Q}(0,s)=0$ and $\lim_{x \to \pm \infty} \tilde{Q}(x,s)<\infty$.
First we take out the homogeneous part by redefining:
$y(x,s)=\tilde{Q}(x,s)-\frac{1}{s}$. This gives us
\bea
\frac{d^2 y}{dx^2}-\gamma_R x \frac{dy}{dx}-\frac{s}{D_R}y=0~,
\eea
where $\gamma_R=k/D_R$.
Let's now do the following change of variable:
\bea
x=\sqrt{\frac{\xi}{\gamma_R}}~,~~~y(x(\xi))=w(\xi)~.
\eea
This gives us
\bea
\xi \frac{d^2 w}{d\xi^2}+\frac{1}{2}(1-\xi)\frac{dw}{d\xi}-aw=0~~,~~~a=\frac{s}{4k}~.
\eea
Getting solutions for the above equation from \cite{ref-1,ref-2} and reverting back to $y(x,s)$ we finally arrive at
\bea
y(x,s)=c_1 x \sqrt{\gamma_R} ~\mathcal{U}\left(\frac{1+4a}{2},\frac{3}{2},\frac{\gamma_R x^2}{2}\right)~,~x>0~,
\eea
where we have imposed the boundary condition $y(x \to \infty,s) < \infty$. Note that here $\mathcal{U}(a,b,z)$ is the
Tricomi confluent hypergeometric function  \cite{ref-1,ref-2}.
Further noting that $y(0,s)=-1/s$, we find
\bea
c_1~ \frac{\sqrt{2\pi}}{\Gamma(\frac{1}{2}+2a)}=-1/s,~~x>0~,
\eea
where $\Gamma(z)$ is the Gamma function.
Thus the solution for the survival probability in the positive $x$-axis reads
\bea
\tilde{Q}^+(x,s)=\frac{1}{s} \left[ 1-\frac{\Gamma(\frac{1}{2}+2a)}{\sqrt{2\pi}} x \sqrt{\gamma_R}~\mathcal{U}\left(\frac{1+4a}{2},\frac{3}{2},\frac{\gamma_R  x^2}{2}\right)  \right].\nonumber \\\hspace{-2.5cm}
\eea
Since the solution is symmetric around zero, we have $\tilde{Q}^-(x,s)=\tilde{Q}^+(-x,s)$. Using the above expressions for the survival probability and substituting in \eref{MFPT-Main}, we arrive at the following expression for the mean first passage time
\bea
\overline{\tau}(x) =-\frac{\mathcal{U}^{(1,0,0)}\left(0,\frac{1}{2},\frac{\gamma_R  x^2}{2}\right)+ \psi^{(0)}\left(\frac{1}{2}\right)}{2D_R \gamma_R  }>0,\hspace{0.5cm}
   \label{MFPT-H2}
\eea
where  $\mathcal{U}^{(1,0,0)}(a,b,z)$ is the partial derivative of the Tricomi confluent hypergeometric function $\mathcal{U}(a,b,z)$ with respect to $a$ and $\psi^{(0)}(z)$ is the digamma function defined as 
the logarithmic derivative of the gamma function i.e., $\psi^{(0)}(z)=\Gamma'(z)/\Gamma(z)$ \cite{ref-1,ref-2}. In particular, one finds $\psi^{(0)}(\frac{1}{2})=-(\gamma+\ln{4})=-1.9635$, where $\gamma$ is the Euler's constant. Substituting this into \eref{MFPT-H2} gives us the solution for the mean first passage time to the origin (i.e., \eref{tau-H}) for a Brownian particle diffusing in a harmonic potential given that it had started from $x$. Now averaging $x$ over the steady state density $\rho_{\text{inst}}(x)$, we find the mean return time $\langle  \overline{\tau}(x)  \rangle=\int~dx~\overline{\tau}(x)\rho_{\text{inst}}(x)$ (recall \eref{taux}). Note that, the latter integral can not be computed exactly. Hence we evaluate the integral numerically.\\

\section{Method of numerical simulation}
\label{simulation}
In the following, we outline the method
that is used to numerically simulate our set up (see Sec. \ref{sec:Setup}). In particular, we consider that in phase-I, the particle performs a simple diffusion. Upon resetting, the particle returns under the action of force $F(x)$ generated by the potential $U(x)$ centered at the resetting location.
To obtain the distribution of the position of the particle at time $t$, we discretize the time $t=n~dt$, where $n$ is an integer. We initialize the position of the particle in the diffusion phase $x(0)=0$, and then, for the first interval $i=1$, the particle evolves under the following discrete version of the Langevin dynamics
\begin{align}
    x(dt)=x(0)+\sqrt{2~D~dt}~\eta,
\end{align}
where $\eta$ is the Gaussian random variable with mean $0$ and variance $1$, and $dt$ is the microscopic time step. The evolution in the remaining steps is as follows. In the next step ($i=2$), the particle can either reset or it can keep evolving. Thus,
\begin{enumerate}
    \item with probability $1-r~dt$ ($r$ is the rate of resetting), the particle still undergoes diffusion so that
\begin{align}
    x(i~dt)=x((i-1)~dt)+\sqrt{2~D~dt}~\eta,
\end{align}
\item whereas with complementary probability $r~dt$, resetting occurs, and the return phase resumes. Herein, the particle diffuses in a potential $U(x)$ centered at the resetting location (the origin). The discreteized version of the dynamics in the return phase then reads
\begin{align}
    x(i~dt)=x((i-1)~dt)-U'[x(i~dt)]dt+\sqrt{2~D_R~dt}~\eta.\label{ret-d}
\end{align}
\end{enumerate}
The dynamics governed by \eref{ret-d} continues until it hits the minimum of the potential \textit{for the first time}. In other words, we stop the return phase dynamics when for the first time $x(t)$ changes sign. The particle now reenters the diffusive phase, and we  proceed as outlined above until $i=n$. We record the position of the particle as a function of observation time for $\mathcal{N}_R$ number of independent realizations.
Finally, we construct the histogram of the position of the particle averaged over these realizations.

\bigskip


\begin{thebibliography}{}



\bibitem{Restart1} Evans, M.R. and Majumdar, S.N., 2011. Diffusion with stochastic resetting. Physical review letters, 106(16), p.160601.

\bibitem{Restart2} Evans, M.R. and Majumdar, S.N., 2011. Diffusion with optimal resetting. Journal of Physics A: Mathematical and Theoretical, 44(43), p.435001.

\bibitem{Kirone} 
Evans, M.R., Majumdar, S.N. and Mallick, K., 2013. Optimal diffusive search: nonequilibrium resetting versus equilibrium dynamics. Journal of Physics A: Mathematical and Theoretical, 46(18), p.185001.


\bibitem{transport1}
Majumdar, S.N., Sabhapandit, S. and Schehr, G., 2015. Dynamical transition in the temporal relaxation of stochastic processes under resetting. Physical Review E, 91(5), p.052131.

\bibitem{transport2}
Eule, S. and Metzger, J.J., 2016. Non-equilibrium steady states of stochastic processes with intermittent resetting. New Journal of Physics, 18(3), p.033006.

\bibitem{Durang}
Durang, X., Henkel, M. and Park, H., 2014. The statistical mechanics of the coagulation–diffusion process with a stochastic reset. Journal of Physics A: Mathematical and Theoretical, 47(4), p.045002.

\bibitem{Pal-potential}
Pal, A., 2015. Diffusion in a potential landscape with stochastic resetting. Physical Review E, 91(1), p.012113.

\bibitem{Pal-time-dep}Pal, A., Kundu, A. and Evans, M.R., 2016.
Diffusion under time-dependent resetting. Journal of Physics A: Mathematical
and Theoretical, 49(22), p.225001.


\bibitem{VV} Pal, A. and Prasad, V.V., 2019. First passage under stochastic resetting in an interval. Physical Review E, 99(3), p.032123.

\bibitem{SRRW} M\'endez, V. and Campos, D., 2016. Characterization of stationary states in random walks with stochastic resetting. Physical Review E, 93(2), p.022106.

\bibitem{underdamped}
Gupta, D., 2019. Stochastic resetting in underdamped Brownian motion. Journal of Statistical Mechanics: Theory and Experiment, 2019(3), p.033212.

\bibitem{local-time}
Pal, A., Chatterjee, R., Reuveni, S. and Kundu, A., 2019. Local time of diffusion with stochastic resetting. Journal of Physics A: Mathematical and Theoretical, 52(26), p.264002.


\bibitem{restart_conc17} Meylahn, J.M., Sabhapandit, S. and Touchette, H., 2015. Large deviations for Markov processes with resetting. Physical Review E, 92(6), p.062148.

\bibitem{magnetic}
Abdoli, I., Vuijk, H.D., Wittmann, R., Sommer, J.U., Brader, J.M. and Sharma, A., 2020. Novel stationary state in Brownian systems with Lorentz force. arXiv preprint arXiv:2004.04471.

\bibitem{ca-sa}
Plata, C.A., Gupta, D. and Azaele, S., 2020. Asymmetric Stochastic Resetting: Modeling Catastrophic Events. arXiv preprint arXiv:2007.12417.


\bibitem{ReuveniEnzyme1}
Reuveni, S., Urbakh, M. and Klafter, J., 2014. Role of substrate unbinding in Michaelis–Menten enzymatic reactions. Proceedings of the National Academy of Sciences, 111(12), pp.4391-4396.


\bibitem{ReuveniEnzyme3}
Rotbart, T., Reuveni, S. and Urbakh, M., 2015. Michaelis-Menten reaction scheme as a unified approach towards the optimal restart problem. Physical Review E, 92(6), p.060101.


\bibitem{bio}
Roldan, E., Lisica, A., Sanchez-Taltavull, D. and Grill, S.W., 2016. Stochastic resetting in backtrack recovery by RNA polymerases. Physical Review E, 93(6), p.062411.


\bibitem{RT}
Evans, M.R. and Majumdar, S.N., 2018. Run and tumble particle under resetting: a renewal approach. Journal of Physics A: Mathematical and Theoretical, 51(47), p.475003.

\bibitem{Luby}
Luby, M., Sinclair, A. and Zuckerman, D., 1993. Optimal speedup of Las Vegas algorithms. Information Processing Letters, 47(4), pp.173-180.


\bibitem{Montanari}
Montanari, A. and Zecchina, R., 2002. Optimizing searches via rare events. Physical review letters, 88(17), p.178701.


\bibitem{Restart-Search1} Kusmierz, L., Majumdar, S.N., Sabhapandit, S. and Schehr, G., 2014. First order transition for the optimal search time of L\'evy flights with resetting. Physical review letters, 113(22), p.220602.

\bibitem{Restart-Search2}
Bressloff, P.C., 2020. Directed intermittent search with stochastic resetting. Journal of Physics A: Mathematical and Theoretical, 53(10), p.105001.

\bibitem{Restart-Search3}
Falcon-Cortes, A., Boyer, D., Giuggioli, L. and Majumdar, S.N., 2017. Localization transition induced by learning in random searches. Physical review letters, 119(14), p.140603.

\bibitem{Chechkin} Chechkin, A. and Sokolov, I.M., 2018. Random search with resetting: a unified renewal approach. Physical review letters, 121(5), p.050601.

\bibitem{HRS}
Pal, A., Ku\'smierz, \L{} and Reuveni, S., 2019. Home-range search provides advantage under high uncertainty. arXiv preprint arXiv:1906.06987.

\bibitem{Pal-19-06}Pal, A., Ku\'smierz, \L{} and Reuveni, S., 2019. Invariants of motion with stochastic resetting and space-time coupled returns, New J. Phys. 21 113024

\bibitem{Review} Evans, M.R., Majumdar, S.N. and Schehr, G., 2020. Stochastic resetting and applications. Journal of Physics A: Mathematical and Theoretical.


\bibitem{expt}
Tal-Friedman, O., Pal, A., Sekhon, A., Reuveni, S. and Roichman, Y., 2020. Experimental realization of diffusion with stochastic resetting. J. Phys. Chem. Lett. 2020, 11, 17, 7350–7355

\bibitem{expt2}
Besga, B., Bovon, A., Petrosyan, A., Majumdar, S. and Ciliberto, S., 2020. Optimal mean first-passage time for a Brownian searcher subjected to resetting: experimental and theoretical results. Phys. Rev. Research 2, 032029(R) (2020).

\bibitem{anamolous-1}
Masoliver, J. and Montero, M., 2019. Anomalous diffusion under stochastic resettings: A general approach. Physical Review E, 100(4), p.042103.


\bibitem{anamolous-3}Maso-Puigdellosas, A., Campos, D. and Mendez, V., 2019. Anomalous diffusion in random-walks with memory-induced relocations. Frontiers in Physics, 7, p.112.

\bibitem{FBM}
dos Santos, M.A., 2019. Fractional Prabhakar derivative in diffusion equation with non-static stochastic resetting. Physics, 1(1), pp.40-58.

\bibitem{scaled}
Bodrova, A.S., Chechkin, A.V. and Sokolov, I.M., 2019. Nonrenewal resetting of scaled Brownian motion. Physical Review E, 100(1), p.012119.

\bibitem{CTRW1}
Shkilev, V.P., 2017. Continuous-time random walk under time-dependent resetting. Physical Review E, 96(1), p.012126.

\bibitem{CTRW2}
Kusmierz, L.. and Gudowska-Nowak, E., 2019. Subdiffusive continuous-time random walks with stochastic resetting. Physical Review E, 99(5), p.052116.

\bibitem{Levy-flight}
Kusmierz, L. and Gudowska-Nowak, E., 2015. Optimal first-arrival times in L\'evy flights with resetting. Physical Review E, 92(5), p.052127.


\bibitem{KPZ}
Gupta, S., Majumdar, S.N. and Schehr, G., 2014. Fluctuating interfaces subject to stochastic resetting. Physical review letters, 112(22), p.220601.

\bibitem{SEP}Basu, U., Kundu, A. and Pal, A., 2019. Symmetric exclusion process under stochastic resetting. Physical Review E, 100(3), p.032136.


\bibitem{ASEP}
Karthika, S. and Nagar, A., 2020. Totally asymmetric simple exclusion process with resetting. Journal of Physics A: Mathematical and Theoretical, 53(11), p.115003.

\bibitem{SEP2} Sadekar, O. and Basu, U., 2020. Zero-current Nonequilibrium State in Symmetric Exclusion Process with Dichotomous Stochastic Resetting. J. Stat. Mech. 073209.


\bibitem{ReuveniPRL}Reuveni, S., 2016. Optimal stochastic restart renders fluctuations in first passage times universal. Physical review letters, 116(17), p.170601.

\bibitem{PalReuveniPRL} Pal, A. and Reuveni, S., 2017. First Passage under Restart. Physical review letters, 118(3), p.030603.

\bibitem{branching}
Pal, A., Eliazar, I. and Reuveni, S., 2019. First passage under restart with branching. Physical review letters, 122(2), p.020602.


\bibitem{Belan} Belan, S., 2018. Restart could optimize the probability of success in a Bernoulli trial. Physical review letters, 120(8), p.080601.

\bibitem{Landau}
Pal, A. and Prasad, V.V., 2019. Landau-like expansion for phase transitions in stochastic resetting. Physical Review Research, 1(3), p.032001.


\bibitem{Peclet}
Ray, S., Mondal, D. and Reuveni, S., 2019. Peclet number governs transition to acceleratory restart in drift-diffusion. Journal of Physics A: Mathematical and Theoretical, 52(25), p.255002.

\bibitem{Gupta}
Nagar, A. and Gupta, S., 2016. Diffusion with stochastic resetting at power-law times. Physical Review E, 93(6), p.060102.

\bibitem{log}Ray, S. and Reuveni, S., 2020. Diffusion with resetting in a logarithmic potential. J. Chem. Phys. 152, 234110.

\bibitem{ORR}
Ahmad, S., Nayak, I., Bansal, A., Nandi, A. and Das, D., 2019. First passage of a particle in a potential under stochastic resetting: A vanishing transition of optimal resetting rate. Physical Review E, 99(2), p.022130.


\bibitem{invariance1}
Pal, A., Ku\'smierz, \L{} and Reuveni, S., 2019. Invariants of motion with stochastic resetting and space-time coupled returns. New Journal of Physics, 21(11), p.113024.



\bibitem{invariance2}
Pal, A., Ku\'smierz, \L{} and Reuveni, S., 2019. Time-dependent density of diffusion with stochastic resetting is invariant to return speed. Physical Review E, 100(4), p.040101.

\bibitem{return3} Bodrova, A.S. and Sokolov, I.M., 2019. Resetting processes with noninstantaneous return. Physical Review E, 101(5), p.052130.

\bibitem{return4} Maso-Puigdellosas, A., Campos, D. and Mendez, V., 2019. Transport properties of random walks under stochastic noninstantaneous resetting. Physical Review E, 100(4), p.042104.

\bibitem{refractory}
Evans, M.R. and Majumdar, S.N., 2018. Effects of refractory period on stochastic resetting. Journal of Physics A: Mathematical and Theoretical, 52(1), p.01LT01.

\bibitem{optical-trap}
Neuman, K.C. and Block, S.M., 2004. Optical trapping. Review of scientific instruments, 75(9), pp.2787-2809.

\bibitem{ratchet-rev}Frank J\"{u}licher, Armand Ajdari, and Jacques Prost, (1997). Modeling molecular motors, Rev. Mod. Phys. 69, 1269.

\bibitem{ref-1}
Zaitsev, V.F. and Polyanin, A.D., 2002. Handbook of exact solutions for ordinary differential equations. CRC press.

\bibitem{ref-2}
Olver, F.W., Lozier, D.W., Boisvert, R.F. and Clark, C.W. eds., 2010. NIST handbook of mathematical functions hardback and CD-ROM. Cambridge university press.





\bibitem{thermo1} Fuchs, J., Goldt, S. and Seifert, U., 2016. Stochastic thermodynamics of resetting. EPL (Europhysics Letters), 113(6), p.60009.

\bibitem{thermo2}
Pal, A. and Rahav, S., 2017. Integral fluctuation theorems for stochastic resetting systems. Physical Review E, 96(6), p.062135.

\bibitem{thermo3}
Gupta, D., Plata, C.A. and Pal, A., 2020. Work fluctuations and Jarzynski equality in stochastic resetting. Physical Review Letters, 124(11), p.110608.

\bibitem{thermo4}
Busiello, D.M., Gupta, D. and Maritan, A., 2020. Entropy production in systems with unidirectional transitions. Physical Review Research, 2(2), p.023011.

\bibitem{thermo5}
Pal, A., Reuveni, S. and Rahav, S., 2020. Thermodynamic uncertainty relation for systems with unidirectional transitions. arXiv preprint arXiv:2008.06953.

\bibitem{extreme}
Majumdar, S.N., Pal, A. and Schehr, G., 2019. Extreme value statistics of correlated random variables: a pedagogical review. Physics Reports, 840, pp.1-32.

\bibitem{intermittent-potential}
Mercado-V\'asquez, G., Boyer, D., Majumdar, S.N. and Schehr, G., 2020. Intermittent resetting potentials. arXiv preprint arXiv:2007.15696.




\bibitem{RednerBook}Redner, S., 2007. A Guide to First-Passage Processes. A Guide to First-Passage Processes, by Sidney Redner, Cambridge, UK: Cambridge University Press, 2007.



\bibitem{Schehr-review}Bray, A.J., Majumdar, S.N. and Schehr, G., 2013. Persistence and first-passage properties in nonequilibrium systems. Advances in Physics, 62(3), pp.225-361.

\bibitem{Redner-review}
Metzler, R., Oshanin, G. and Redner, S. eds., 2014. First-passage phenomena and their applications. World Scientific Publishing.

\bibitem{Grebenkov}
Grebenkov, D.S., 2014. First exit times of harmonically trapped particles: a didactic review. Journal of Physics A: Mathematical and Theoretical, 48(1), p.013001.

\bibitem{Benichou}
Benichou, O., Loverdo, C., Moreau, M. and Voituriez, R., 2011. Intermittent search strategies. Reviews of Modern Physics, 83(1), p.81.



\end{thebibliography}

\end{document}